\documentclass[a4paper,10pt]{edpsci}

\usepackage{amsmath,amssymb,amsfonts}
\usepackage{graphicx}
\usepackage{epstopdf}
\usepackage[numbers]{natbib}
\usepackage{hyperref}
\usepackage{url}
\usepackage[T1]{fontenc}
\usepackage[utf8]{inputenc}
\usepackage{siunitx}
\usepackage{tabularx}

\usepackage[switch]{lineno}
\hypersetup{colorlinks=true,citecolor=blue,urlcolor=blue,linkcolor=blue}
\journalname{Acta Acustica}

\articletype{Scientific Article}

\begin{document}

\title{Photogrammetry-Reconstructed 3D Head Meshes for Accessible Individual Head-Related Transfer Functions}

%%% Running title for page header
\titlerunning{Acta Acustica}

\author{Ludovic Pirard\inst{1}\correspondingauthor{\email{l.pirard@imperial.ac.uk}} 
\and
Lorenzo Picinali\inst{1}
\and
Katarina C. Poole\inst{1}}

%%% Running author for page header
\authorrunning{L. Pirard et al.}

\institute{Dyson School of Design Engineering, Imperial College London, United Kingdom}

\abstract{Individual head-related transfer functions (HRTFs) are essential for accurate spatial audio binaural rendering but remain difficult to obtain due to measurement complexity. This study investigates whether photogrammetry-reconstructed (PR) head and ear meshes, acquired with consumer hardware, can provide a practically useful baseline for individual HRTF synthesis. Using the SONICOM HRTF dataset, 72-image photogrammetry captures per subject were processed with Apple's Object Capture API to generate PR meshes for 150 subjects. Mesh2HRTF was used to compute PR synthetic HRTFs, which were compared against measured HRTFs, high-resolution 3D scan-derived HRTFs, KEMAR, and random HRTFs through numerical evaluation, auditory models, and a behavioural sound localisation experiment (N = 27). PR synthetic HRTFs preserved ITD cues but exhibited increased ILD and spectral errors. Auditory-model predictions and behavioural data showed substantially higher quadrant error rates, reduced elevation accuracy, and greater front–back confusions than measured HRTFs, performing worse than random HRTFs on perceptual metrics. Current photogrammetry pipelines support individual HRTF synthesis but are limited by insufficient pinna morphology details and high-frequency spectral fidelity needed for accurate individual HRTFs containing monaural cues.}

\keywords{HRTF, Photogrammetry Reconstruction, Mesh2HRTF, Spatial audio, Binaural rendering}

\maketitle
\section{Introduction}
Spatial hearing, the ability to identify the direction of sounds in three-dimensional space, is crucial for everyday listening. It relies on acoustic cues generated by the interaction of sound waves with the listener's head, torso, and ears. Binaural cues, including interaural time differences (ITDs) and interaural level differences (ILDs), are particularly effective for lateralisation in the horizontal plane \cite{middlebrooks_sound_1991}. However, accurate elevation perception and front–back discrimination additionally depend on monaural spectral cues generated by the complex geometry of the pinnae, head, and torso, which shape the sound spectrum in a direction-dependent manner \cite{langendijk_contribution_2002}.

Head-related transfer functions (HRTFs) describe these direction-dependent acoustic filtering properties, characterising how sound from a given location is transformed before reaching the entrance of the ear canal \cite{engelSONICOMHRTFDataset2023}. Generic HRTFs derived from average human morphology, such as those measured using a KEMAR acoustic mannequin \cite{gardner_hrtf_1995}, are widely used in consumer applications but often fail to reproduce individual spectral cues, leading to front–back confusions and impaired elevation perception \cite{brinkmann_cross-evaluated_2019}. Individual HRTFs are therefore fundamental for high-quality spatial audio, improving localisation accuracy and spatial release from masking \cite{gonzalez-toledo_spatial_2024, picinali_system--user_2023, meyer_generalization_2025}.

Traditional acquisition of individual HRTFs requires anechoic facilities, loudspeaker arrays, in-ear microphones, and specialist knowledge \cite{pauwels_relevance_2023}. Numerical synthesis from high-resolution 3D head and ear meshes using tools such as Mesh2HRTF \cite{ziegelwanger_mesh2hrtf_2015} offers a more accessible alternative, but access to laser scans or structured-light scanners remains limited due to cost and logistical constraints \cite{pollack_perspective_2022, ziegelwanger_calculation_2013}. Consequently, many users still rely on generic or poorly matched HRTFs, and scalable solutions for individual cues remain an open challenge.

A wide range of HRTF individualisation strategies has been explored to reduce acquisition burden. Anthropometric approaches relate physical measurements of the head and torso to HRTF characteristics and select or morph HRTFs from a database \cite{spagnol_hrtf_2020}, but require precise measurements at defined landmarks, which can be time-consuming and error-prone in practice. Image-based and machine-learning methods combine photographs of the ear and head with anthropometry \cite{zhao_magnitude_2022, lee_personalized_2018}, or attempt to extract geometric features directly from 3D meshes \cite{fantini_hrtf_2021, manoj_extraction_2016}. Whilst these approaches reduce explicit measurement effort, they still rely on controlled acquisition conditions or specialised scanning equipment, limiting their accessibility.

More recently, data-driven techniques have sought to reconstruct full HRTF sets from sparse acoustic measurements \cite{hogg_hrtf_2024, hu_hrtf_2024, hu_machine_2025}. These methods can substantially reduce measurement time but still require dedicated loudspeaker setups and recording sessions, and are therefore difficult to deploy at scale outside research environments. Overall, existing individualisation methods tend to trade off between accuracy, cost, and user effort, and none yet offers a widely deployable path to individual HRTFs for the general public.

Photogrammetry presents a more accessible alternative for 3D head and ear capture, as it uses conventional cameras or smartphones to capture surface geometry from multiple overlapping images. Prior work has investigated photogrammetry-based HRTFs \cite{pollack_perspective_2022, dellepiane_reconstructing_2008, meshram_p-hrtf_2014}, but limited mesh resolution and insufficient ear detail have so far constrained performance, particularly at high frequencies where fine pinna morphology is critical \cite{pollack_application_2023, pollack_combination_2024}. It therefore remains unclear to what extent photogrammetry, implemented with current consumer hardware and reconstruction pipelines, can provide a practically useful basis for individual HRTF synthesis.

In this study, head and ear meshes reconstructed using photogrammetry data from the SONICOM dataset are used as input to Mesh2HRTF in order to synthesise individual HRTFs for 150 subjects. The resulting photogrammetry-reconstructed (PR) synthetic HRTFs are compared against acoustically measured HRTFs, HRTFs computed from high-resolution 3D scans \cite{poole_extended_2025}, a measured KEMAR reference, and randomly selected measured HRTFs within the SONICOM dataset. Evaluation combines numerical metrics targeting key localisation cues (ITD, ILD, and log-spectral distortion for monaural cues), behavioural predictions from two auditory spatial perception models, and a virtual-reality sound localisation experiment.

The aim of this work is not to introduce a new individualisation algorithm, but to assess whether photogrammetry-based reconstruction, with minimal subject input data and commodity hardware, can provide a practically useful starting point for individual HRTFs. Specifically, this study addresses three research questions. RQ1 - do PR synthetic HRTFs preserve the key numerical characteristics of measured HRTFs, particularly interaural time and level differences and monaural spectral cues to a similar extent as 3D synthetic HRTFs computed from high-resolution scans? RQ2 - do any observed numerical differences translate into predicted localisation performance differences when assessed using computational auditory models? RQ3 - how do PR synthetic HRTFs perform in actual sound localisation tasks with human listeners compared to individual measured and randomly selected non-individual HRTFs? By addressing these questions, the study delineates the current capabilities and limitations of this accessible approach and identifies priorities for future improvements in reconstruction, processing, and refinement.

\section{Methods}

\subsection{Data Acquisition}

This study uses pre-existing photogrammetry data from the SONICOM HRTF dataset \cite{engelSONICOMHRTFDataset2023} to assess the feasibility of computing individual HRTFs from photographs. For each of the 150 selected subjects, 72 images had been captured at \(5^{\circ}\) intervals to achieve a full \(360^{\circ}\) representation. The acquisition protocol involved a custom app on an iPhone XS sending Open Sound Control (OSC) messages to control a motorised turntable whilst the iPhone captured photographs as the turntable rotated. The subject was seated on a chair on top of the turntable, and the iPhone was positioned at head height using a tripod, orthogonally aligned to the subject's right side. The distance between the iPhone and the subject varied between sessions.

\begin{figure}[htbp]
    \centering
    \includegraphics[width=0.9\columnwidth]{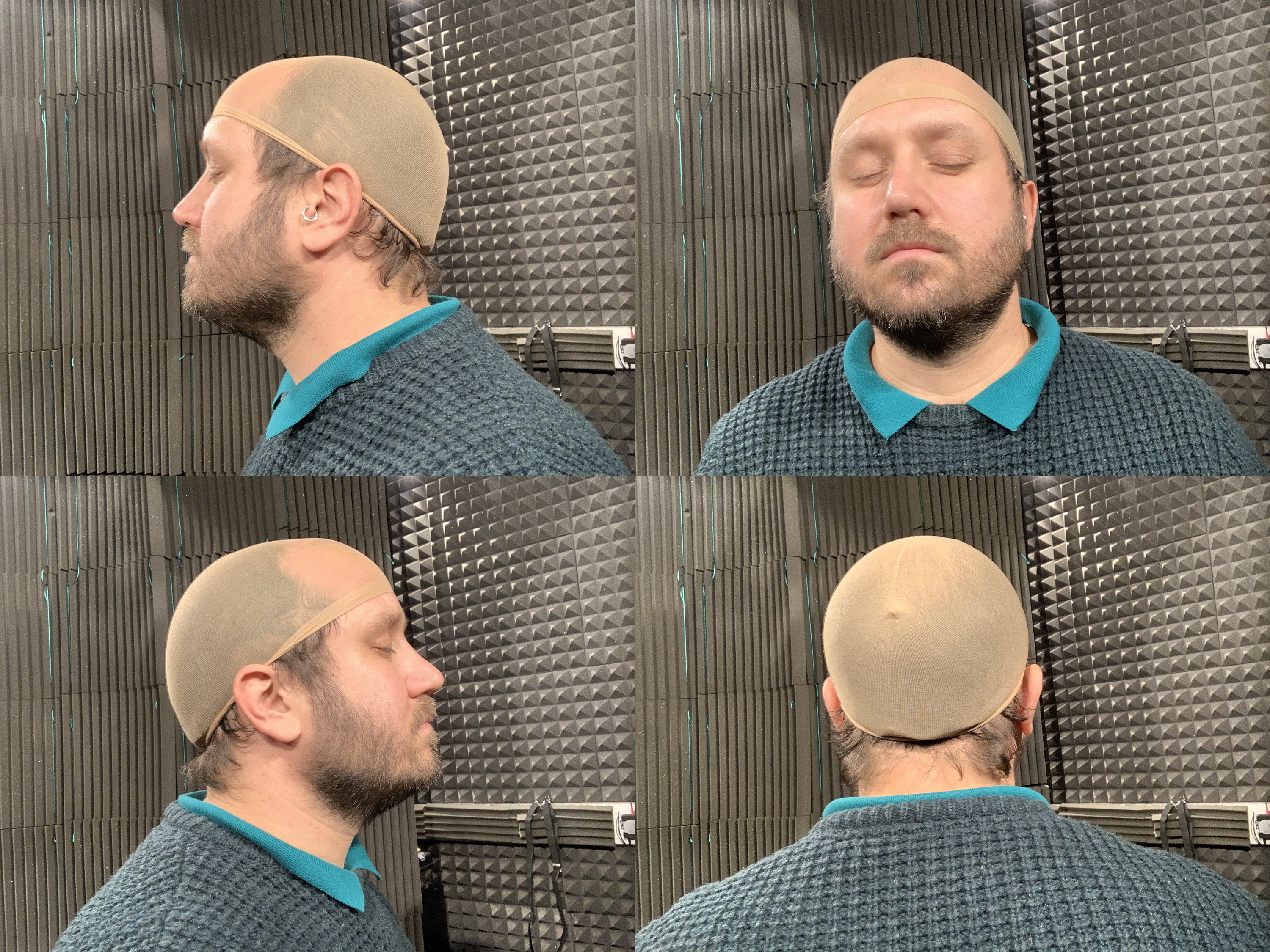}
    \caption{Photogrammetry data example from the SONICOM dataset.}
    \label{fig:PhotoData}
\end{figure}

The pre-existing photogrammetry data include high-resolution RGB photographs (2316\(\times\)3088 pixels), depth maps, and iPhone inertial sensor data. All data were acquired using consumer hardware (iPhone XS) with a custom 3D‑printed mirror bracket to utilise Apple's TrueDepth technology. This technology projects infrared dots and captures their reflections to create depth maps. The photographs are encoded in the HEIC format, which stores the RGB image and corresponding depth map for each file. Figure \ref{fig:PhotoData} shows examples of photogrammetry data for one subject at different angles. 

\subsection{Photogrammetry Reconstruction}

Photogrammetry reconstruction (PR) transforms two-dimensional photographs into a three-dimensional model using structure-from-motion techniques. Common features across overlapping images are identified to compute camera positions and orientations, producing a sparse point cloud that is subsequently densified. From this dense point cloud, a triangulated surface mesh is generated.

Several software packages were evaluated, including Reality Capture, Agisoft Metashape, Autodesk Recap Photo, Meshroom (AliceVision), and Apple's Object Capture API. Most produced reconstructions of poor quality, characterised by noise and artefacts in two key regions, the torso and the upper head, frequently resulting in implausible geometry such as fragmented cranial structures and distorted torso shapes.

An informal evaluation of mesh fidelity, resolution, and ear morphology reveals that Apple's Object Capture API yields the most accurate and visually consistent results. Reconstructions are free from major artefacts, exhibit clean isolation of the subject from the background, and show correct geometry for both the torso and the upper head.

Based on these findings, Apple's Object Capture API is selected for implementation. A batch processing pipeline is developed in Swift using Xcode, with 72 input images per subject. The pipeline outputs a polygonal surface mesh in STL format, representing subject geometry reconstructed from the SONICOM photogrammetry dataset. Each reconstructed mesh is paired with a corresponding high-resolution reference scan acquired with an EXScan Pro from the SONICOM dataset. These reference scans were processed by Poole et al. (2025) \cite{poole_extended_2025}, including scaling, rotation, beheading, surface smoothing, and compatibility checks for Mesh2HRTF. 

\subsection{Mesh Processing}

Before acoustic simulation, all meshes underwent a processing pipeline to ensure geometric consistency and numerical robustness. The photogrammetry-reconstructed (PR) meshes were aligned to their corresponding high-resolution reference scans (which are aligned to the frankfurt plane) using landmark-based registration, followed by iterative closest-point refinement. Meshes were then ``beheaded'' above the shoulders to remove the torso geometry whilst preserving the head and ears for HRTF synthesis.

Residual artefacts, such as isolated components, self-intersections, and non-manifold elements, were removed using a combination of automated filters and manual inspection. Curvature-adaptive mesh grading \cite{palm_curvature-adaptive_2021} was applied to refine the mesh resolution in high-curvature regions, such as the pinnae, whilst coarsening flatter areas (for example, cheeks and scalp). Finally, distinct mesh regions were labelled for the skin, right ear canal, and left ear canal to facilitate material assignment and receiver placement in subsequent simulations.

A minority of reconstructed meshes exhibited excessive geometric complexity due to prominent hair or facial hair, rendering post-processing impractical within computational constraints. Such cases were excluded from subsequent HRTF synthesis.

\subsection{HRTF Synthesis}

Mesh2HRTF is an open-source package that computes listener-specific HRTFs in the standardised SOFA format \cite{majdak_spatially_2013} numerically using a Burton–Miller boundary element method coupled with a multilevel fast multipole method \cite{ziegelwanger_mesh2hrtf_2015}.

Mesh2HRTF was used to compute synthetic HRTFs from PR meshes following the same configuration and methodology used by Poole et al. (2025) \cite{poole_extended_2025} to compute synthetic HRTFs from high-resolution 3D scans. The simulated frequency range was from 0 to 24 kHz in 150 Hz steps and acoustic propagation was solved using ML-FMM BEM solver. Source and receiver positions were matched to the SONICOM HRTF measurement setup to enable direct comparison with acoustically measured HRTFs. The HRTF synthesis process followed the guidelines provided in the Mesh2HRTF documentation to ensure methodological consistency across subjects and conditions. 

Synthetic HRTFs were post-processed through spatial alignment, temporal windowing, level normalisation, and interaural time-difference (ITD) removal. Source positions were first aligned to match the synthetic, measured, and KEMAR reference HRTFs. Each head-related impulse response (HRIR) was then windowed to the length of the KEMAR no-ITD HRTF \footnote{\texttt{"KEMAR\_Knowl\_EarSim\_LargeEars\_Windowed\_NoITD\_48kHz.sofa"}} using a 16-sample sine-squared fade-in and a 128-sample cosine-squared fade-out, preserving early energy while attenuating late reflections. Broadband level was normalised by scaling the HRIRs such that the mean root-mean-square (RMS) level at the frontal direction (\(0^{\circ}\) elevation, \(0^{\circ}\) azimuth) matched the KEMAR reference, ensuring consistent loudness across HRTFs. ITDs were subsequently removed by estimating ear onsets using a threshold-based detector and time-shifting the HRIRs to equalise arrival times, with a fixed \SI{0.8}{ms} padding. Any wrapped samples resulting from the shift were zeroed. The number of shifted samples was stored as metadata in the SOFA file in order to be used later in a binaural renderer. The processed HRIRs were stored in SOFA format \cite{majdak_spatially_2013} as windowed, no-ITD, level-normalised HRTFs.

\subsection{Numerical Assessment}

Synthetic HRTFs obtained from PR meshes were evaluated numerically against acoustically measured HRTFs, synthetic HRTFs computed by Poole et al. (2025) \cite{poole_extended_2025} from high-resolution 3D scans (3D synthetic), a measured KEMAR HRTF, and randomly selected measured HRTFs within the SONICOM dataset. The numerical evaluation focused on three primary localisation cues: the interaural time difference (ITD), the interaural level difference (ILD), and monaural spectral cues. To quantify differences in these monaural cues, the log-spectral distortion (LSD) metric \cite{hu_hrtf_2024} was applied, comparing the frequency spectra of different HRTFs at each source location and for each ear. All numerical analyses were performed using the Spatial Audio Metrics toolbox (version 0.1.2, K. C. Poole, AXD, Imperial College London) \cite{poole_extended_2025}: \url{https://github.com/Katarina-Poole/Spatial-Audio-Metrics}.

Descriptive statistics are reported as the median together with the interquartile range (25th–75th percentile). For each metric, data distributions are first assessed for normality; when the normality assumption holds, repeated-measures ANOVA is used, with Tukey’s honestly significant difference (HSD) post-hoc tests for pairwise comparisons. For metrics with non-normal distributions, the Friedman test is used instead, followed by pairwise Wilcoxon signed-rank tests with  Holm correction for multiple comparisons. For spatial and frequency-dependent analyses, cluster-based permutation testing is applied, with false discovery rate (FDR; Benjamini–Hochberg) correction across clusters to identify regions with significant differences.

\subsubsection{ITD and ILD}

Interaural time differences (ITDs) are evaluated after spatial alignment of each candidate HRTF with the measured HRTF by matching source directions. For each matched direction, ITD is estimated from the left–right HRIR pair using a threshold-based method, yielding the interaural arrival-time difference in microseconds. The per-direction ITD error is defined as the absolute difference between candidate and measured ITDs, and the subject-level ITD metric is the mean absolute ITD error across all directions.

Interaural level differences (ILDs) are computed per direction as the broadband energy ratio between right and left ear HRIRs in decibels, with a small constant $\epsilon$ to avoid taking the log of zero. The per-direction ILD error is the absolute difference between candidate and measured ILDs, and the subject-level ILD metric is the mean absolute ILD error across directions:

\begin{align}
E_L(d) &= \frac{1}{N}\sum_{n=1}^{N} x_L^2[n;d], \quad
E_R(d) = \frac{1}{N}\sum_{n=1}^{N} x_R^2[n;d], \nonumber \\
\text{ILD}_{\bullet}(d) &= 10\log_{10}\left(\frac{E_R(d)+\epsilon}{E_L(d)+\epsilon}\right), \nonumber \\
\varepsilon_{\text{ILD}}(d) &= \big|\text{ILD}_{\text{cond}}(d) - \text{ILD}_{\text{meas}}(d)\big|, \nonumber \\
\text{ILD} &= \frac{1}{D}\sum_{d=1}^{D} \varepsilon_{\text{ILD}}(d) \quad [\text{dB}].
\end{align}

\noindent Where: $x_L[n;d], x_R[n;d]$ are the left/right HRIR samples at direction $d$; $N$ is the HRIR length; $E_L, E_R$ are the broadband energies; $\epsilon = 10^{-10}$; $\text{ILD}_{\bullet}(d)$ is the ILD for either condition or measured HRTF ($\bullet \in \{\text{cond},\text{meas}\}$); $\varepsilon_{\text{ILD}}(d)$ is the absolute ILD error for direction $d$; and $D$ is the number of matched directions.

\subsubsection{Log-Spectral Distortion and Spectral Analysis}

Log-spectral distortion (LSD) aims to quantify monaural spectral fidelity. For each ear and matched direction, the HRIR is transformed into a magnitude spectrum via a fast Fourier transform and compared to the corresponding measured spectrum. The per-direction, per-ear LSD is computed as the root-mean-square (RMS) difference of the two log-magnitude spectra across frequency, with a small stabilising constant $\epsilon$ to avoid taking the logarithm of zero:

\begin{align}
\text{LSD}_e(d) &= 
\sqrt{\frac{1}{K}\sum_{k=1}^{K}
\Bigg[
\begin{aligned}
&20\log_{10}\!\big(|H_{\text{cond}}(e,d,f_k)|+\epsilon\big)\\
&-\,20\log_{10}\!\big(|H_{\text{meas}}(e,d,f_k)|+\epsilon\big)
\end{aligned}
\Bigg]^2}, \nonumber \\
\text{LSD} &= 
\frac{1}{2D}\sum_{d=1}^{D}\sum_{e\in\{L,R\}}
\text{LSD}_e(d)\;[\text{dB}].
\end{align}

\noindent Where: $H_{\text{cond}}(e,d,f_k), H_{\text{meas}}(e,d,f_k)$ are the condition and measured HRTF magnitude at ear $e$, direction $d$, frequency bin $f_k$; $e \in \{L,R\}$ denotes the ear; $d$ is the matched direction; $D$ is the number of matched directions; $K$ is the number of frequency bins; $\epsilon = 10^{-10}$; and $\text{LSD}_e(d)$ is the RMS difference of log-magnitude spectra for ear $e$ and direction $d$.

\subsubsection{Spatial Distribution Analysis}

The spatial distribution of ITD, ILD, and LSD differences across four comparisons, averaged over 150 subjects was examined. For each azimuth–elevation position with 45° steps and each metric, a one-sample t-test was performed across subjects to test whether the signed mean cue difference between HRTFs conditions and measured HRTFs differed from zero. P-values were corrected for multiple comparisons across positions using the FDR procedure (per condition and metric). This allows detailed examination of where these errors occur spatially and which locations show significant differences.

\subsection{Model-Based Assessment}

Whilst numerical evaluation provides indications of the performance of HRTFs based on localisation cues, these metrics do not directly capture their perceptual impact. Auditory models allow numerical representation of how the human auditory system would interpret specific filters for different sound locations. Spherical sound-localisation models are designed to reflect how a human listener would respond to stimuli during a localisation test, thereby characterising the quality of spatialisation obtained using a given set of HRTFs.

\subsubsection{Model Implementation}

Two auditory models from the Auditory Modelling Toolbox (AMT) were employed to predict localisation performance: Baumgartner2014 \cite{baumgartner_modeling_2014}, which compares internal sound representations with a template to yield probabilistic predictions of polar-angle responses, and Barumerli2023 \cite{barumerli_bayesian_2023}, a Bayesian model that performs optimal information processing of spatial cues. Both models use directional transfer functions (DTFs) derived from HRTFs as inputs, with each subject's measured HRTF serving as the template. Target HRTFs tested include the measured HRTF, 3D synthetic HRTF \cite{poole_extended_2025}, PR synthetic HRTF, KEMAR HRTF, and a randomly selected measured HRTF from the SONICOM dataset. 

Model parameters for baumgartner2014 were adopted from Lladó et al. (2025) \cite{llado_spectral_2025}, who showed that a notch‑region spectral weighting around 8 kHz best accounted for listener performance in a localisation task. The parameters correspond to the median of the individually fitted values for the notch‑region model variant across listeners. For barumerli2023, parameters were taken from Daugintis et al. (2023) \cite{daugintis_classifying_2023, daugintis_initial_2023} as the median values fitted across 16 subjects in perceptual localisation experiments. The parameter sets for both models are listed in Table \ref{tab:ModelParams}. Model predictions are statistically compared using repeated measures ANOVA with Tukey's HSD post-hoc tests for pairwise comparisons.

\begin{table}[htbp]
\centering
\caption{Auditory model parameters.}
\label{tab:ModelParams}
\small
\begin{tabularx}{\columnwidth}{@{}llX@{}}
\hline
\textbf{Model} & \textbf{Parameter} & \textbf{Value} \\
\hline
Baumgartner2014 & Sensitivity ($\gamma$) & 3.87 \\
 & Specificity (S) & -6.77 \\
 & Mistake rate ($\epsilon$) & 13.11 \\
 & Max. frequency & 17 kHz \\
 & Lateral angles & $\pm$75°, $\pm$50°, $\pm$25°, 0° \\
\hline
Barumerli2023 & MC simulations ($N_{\mathrm{exp}}$) & 50 \\
 & ILD noise ($\sigma_{\mathrm{ILD}}$) & 0.75 \\
 & Spectral noise ($\sigma_{\mathrm{spectral}}$) & 4.3 \\
 & Prior uncertainty ($\sigma_{\mathrm{prior}}$) & 11.5 \\
 & Motor noise ($\sigma_{\mathrm{motor}}$) & 13.45 \\
\hline
\end{tabularx}
\end{table}

\subsection{Behavioural Assessment}

Whilst auditory models allow numerical representation of how the human auditory system would interpret specific filters for different sound locations, perceptual evaluation through localisation experiments provides direct insight into how synthetic HRTFs perform in listening scenarios. Localisation performance is assessed using the virtual-reality (VR) localisation test developed by Daugintis et al. (2023, 2024) \cite{daugintis_initial_2023, daugintis_effects_2024} in Unity, presented through a Meta Quest 3 VR headset with real-time head and controller tracking. Participants first completed a 87-trial training phase using a loudspeaker array to familiarise themselves with the task. In each trial, a virtual sound source is presented, and participants indicate the perceived direction by pointing with a VR controller and pressing the trigger. The training phase comprised 29 unique source directions, each presented three times in a randomised order, yielding 87 trials in total. For the first 29 trials, the audio source position was indicated visually. Thereafter, the visual indicator was removed and participants were informed that subsequent localisations must rely solely on auditory cues. Colour-coded angular error feedback was provided after each response: green ($\leq$ 5°), yellow ($\leq$ 20°), orange ($\leq$ 35°), and red ($\leq$ 35°). When the angular error exceeded 5°, up to two additional correction attempts were permitted for that position, with the visual source indicator reinstated on the second correction attempt.

Three HRTF conditions are evaluated: (1) the participant's acoustically measured HRTF, (2) the PR synthetic HRTF, and (3) a randomly selected (different for each participant) measured HRTF within the SONICOM dataset. The latter condition was chosen in order to be the benchmark for a non-individual HRTF. Condition order was randomised for each participant and each trial. The stimulus is a sequence of three consecutive Gaussian noise bursts (100 ms each) windowed with a Hann function, resulting in a total duration of 300 ms. The stimulus is presented using the 3D Tune-In Toolkit \cite{3dti_2019} via its Unity Wrapper. Sennheiser HD599 headphones are directly plugged into the VR headset with the same set volume across participants of 65 dBA which varied slightly depending on stimulus position.

The experiment comprises three blocks of 99 trials, covering 33 distinct source positions distributed across 16 azimuths spanning the full horizontal circle (0° to 330°) and elevations from -30° to 90°. Elevation sampling is densest along the cardinal axes (0°, 90°, 180°, 270°), with up to five elevations per azimuth, and includes diagonal directions (45°, 135°, 225°, 315°) at two elevations (-30°, 30°). Intermediate azimuths (30°, 60°, etc.) are sampled at the horizontal plane only. Source positions are presented in a random order for each participant and trial. The pointing direction is recorded together with the corresponding HRTF condition and source position.

\subsection{Behavioural analysis}

Behavioural performance is quantified using seven error metrics: the great circle error (overall angular deviation between target and response on the sphere), the absolute lateral accuracy, the lateral precision, the absolute polar accuracy, the polar precision, the front-back confusion rate and the quadrant error. Front-back confusions are classified using the method of Poirier-Quinot et al. (2022) \cite{poirier_quinot_spatial_2022}, defined as responses falling within a 45\(^\circ\) cone around the mirrored front–back position of the target. Quadrant error is computed in lateral–polar coordinates following Middlebrooks \cite{middlebrooks_virtual_1999} as the percentage of responses that are assigned to a different lateral–polar quadrant than the target.

For each participant, per-HRTF/location performance is aggregated using the median across trials. Group-level statistics are computed as the median together with the interquartile range (25th–75th percentile) of participant medians. Statistical analysis uses the Shapiro-Wilk test to assess normality of the distribution for each metric. For metrics with normal distributions, repeated-measures analysis of variance (ANOVA) is performed, followed by Tukey's HSD post-hoc test for pairwise comparisons. For metrics with non-normal distributions, the Friedman test is used, followed by pairwise Wilcoxon signed-rank tests with Holm correction for multiple comparisons. Pearson correlation coefficients are computed to quantify the linear relationship between target and response positions in horizontal and median planes.

To examine whether spectral similarity between measured and randomly selected HRTFs influences localisation performance, a supplementary correlation analysis was conducted. For each of the 27 participants, the log-spectral distortion (LSD) between their measured HRTF and the assigned random HRTF was computed over the 1–16~kHz frequency range, averaged across all spatial directions and both ears. Pearson correlation coefficients were then calculated between these LSD values and the performance differences between non-individual and individual (Random minus Measured) for four behavioural metrics.

\section{Results}

\subsection{3D Reconstruction}

To describe qualitatively how photogrammetry-reconstructed (PR) meshes compare to high-resolution 3D scans, wireframe views were inspected in Blender for all subjects. This comparison is descriptive only; numerical geometric metrics were not applied, as differences in mesh source, neck cut, and alignment would confound direct mesh-to-mesh distance measures. The following observations summarise what was seen across the reconstructed meshes.

\begin{figure}[htbp]
    \centering
    \includegraphics[width=0.49\columnwidth]{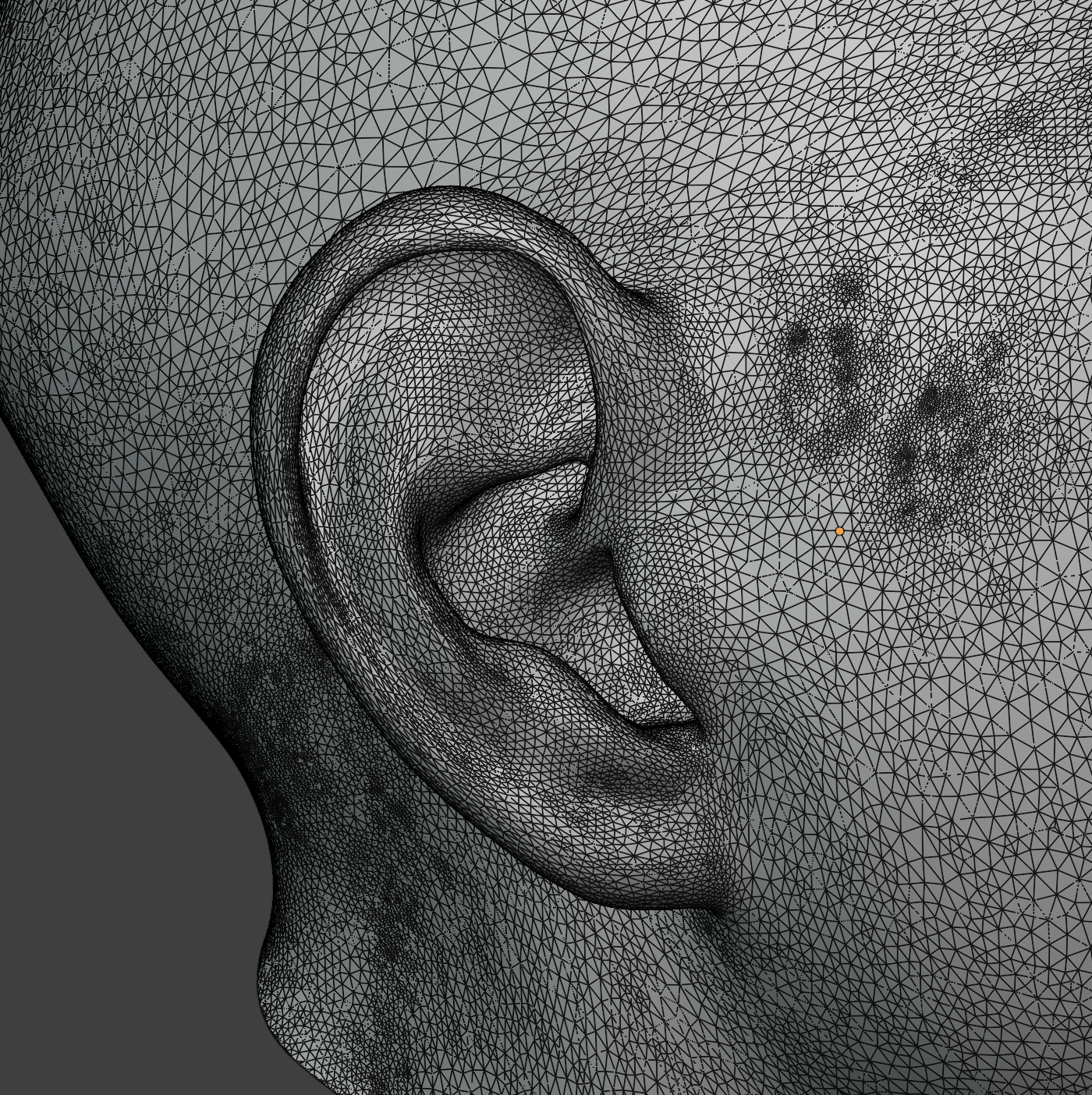}
    \includegraphics[width=0.49\columnwidth]{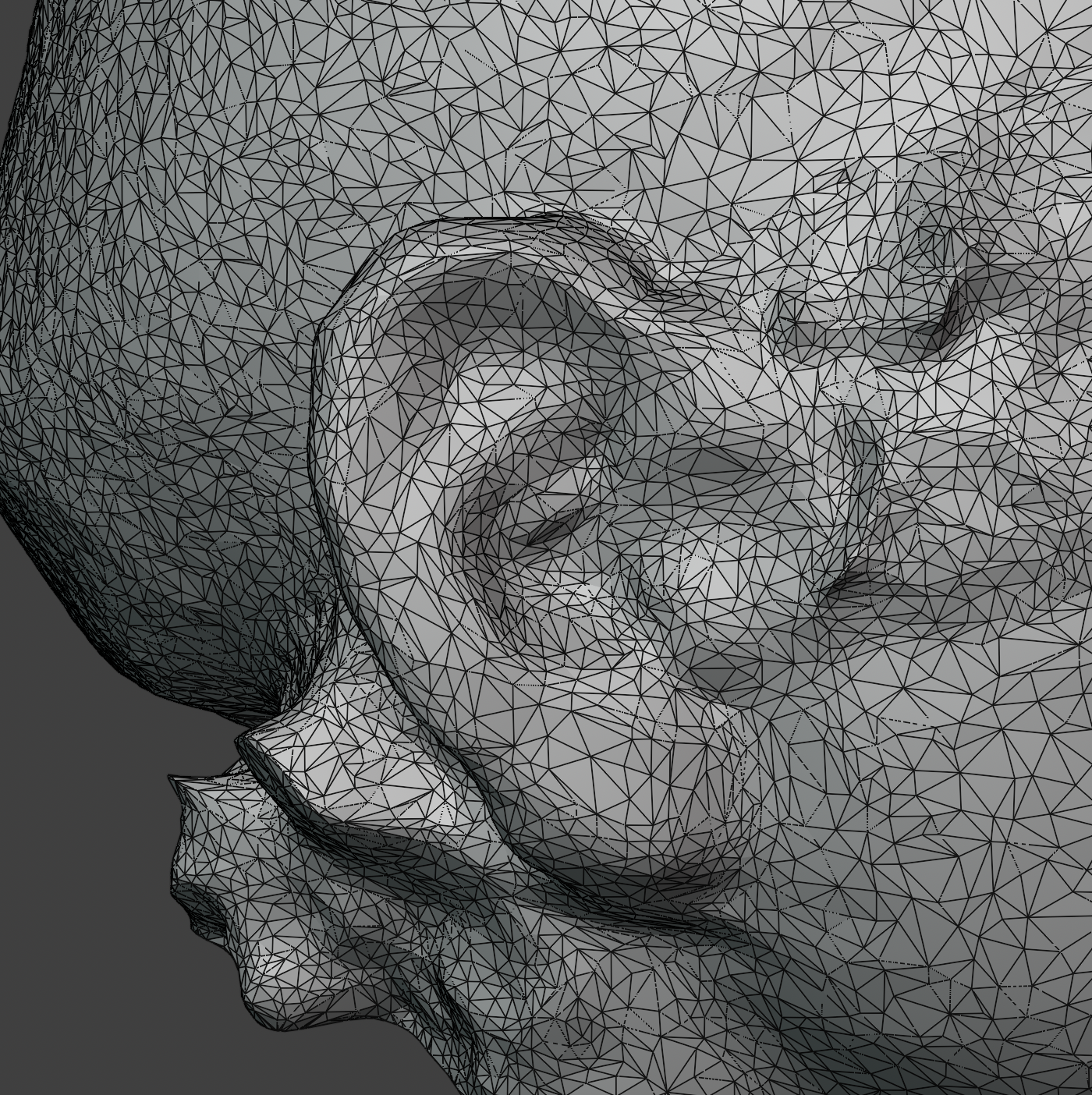}
    \caption{Wireframe views of the high-resolution 3D scan mesh (left, subject P0004, right ear) and the photogrammetry-reconstructed mesh (right, same subject and same ear).}
    \label{fig:MeshesComparison}
\end{figure}

Globally, both reconstructions capture the main features of each subject, including head width, eyes positions, lips, nose, chin, cheeks, and forehead. However, clear discrepancies are observed in the external ear region. The principal differences concern the number of faces and vertices, which determine mesh resolution and the level of morphological detail. The 3D scans, illustrated in Figure \ref{fig:MeshesComparison} (left), provide highly detailed and anatomically accurate representations of the pinna, with well-defined structures such as the helix, lobule, antihelix, and concha. In contrast, the PR meshes, illustrated in Figure \ref{fig:MeshesComparison} (right), exhibit coarse geometry in this region, with oversimplified pinna features that did not capture individual ear morphology. PR meshes were scaled to match the high-resolution reference before HRTF computation. Although fine ear morphology is not reconstructed, the PR meshes retain sufficient overall head and ear geometry for initial HRTF synthesis. 

\subsection{Spectral Characteristics}

\begin{figure}[htbp]
    \centering
    \includegraphics[width=\columnwidth]{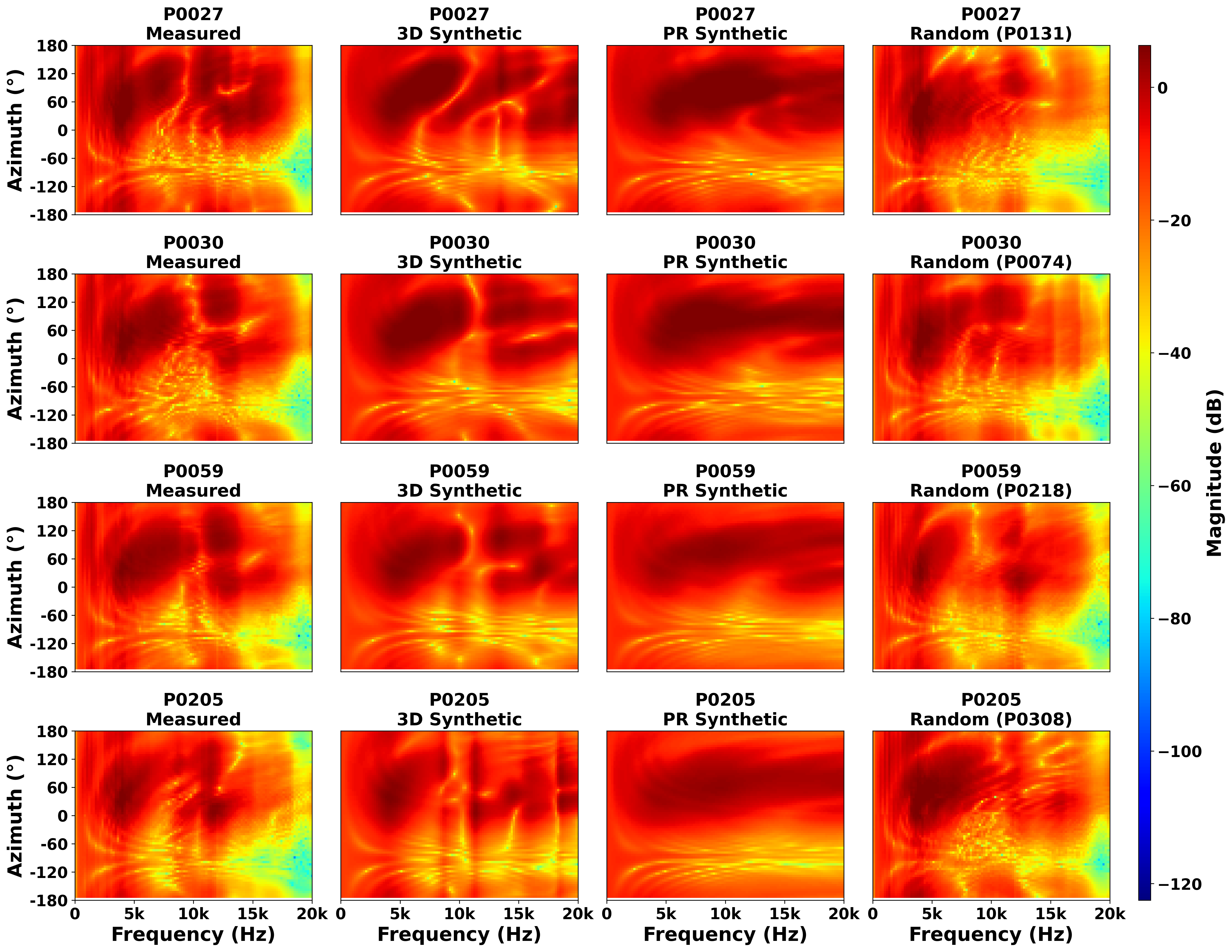}
    \caption{Azimuth plane: Four subjects measured, 3D synthetic, PR synthetic and random HRTFs (left HRTFs).}
    \label{fig:4Azimuth}
\end{figure}

\begin{figure}[htbp]
    \centering
    \includegraphics[width=\columnwidth]{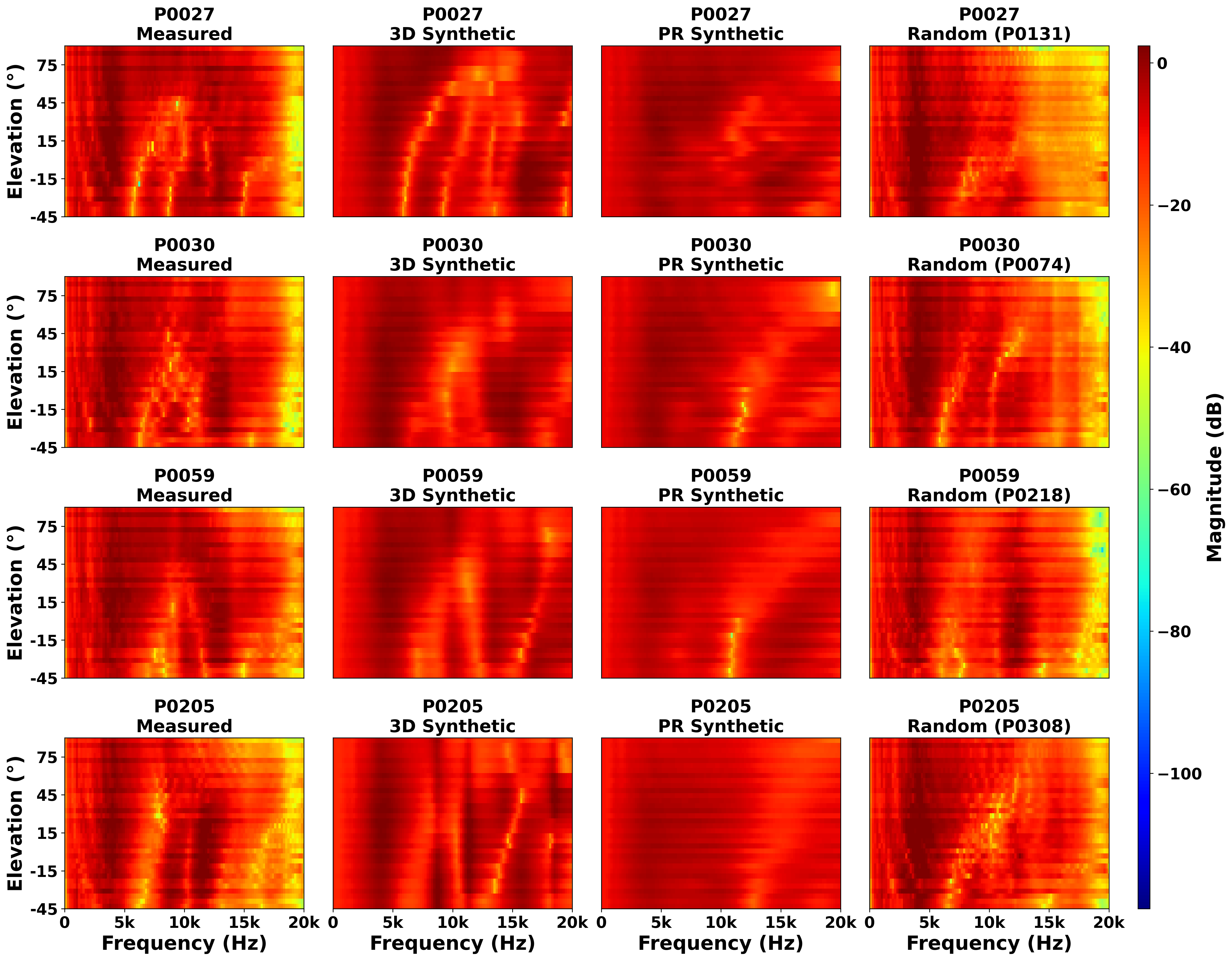}
    \caption{Elevation plane: Four subjects measured, 3D synthetic, PR synthetic and random HRTFs (left HRTFs).}
    \label{fig:4Elevation}
\end{figure}

Spectral representations allow inspection of direction-dependent spectral cues, such as pinna-related peaks and notches, and facilitate qualitative comparison of spectral fine structure between individual and non-individual measured, 3D synthetic, and PR synthetic HRTFs. Magnitude responses are visualised as heatmaps over azimuth and elevation. The comparisons in Figs.~\ref{fig:4Azimuth} and \ref{fig:4Elevation} illustrate averaged magnitude responses (dB) of left ear HRIRs for four randomly selected subjects in the azimuth plane (elevation \(0^{\circ}\)) and elevation plane (azimuth \(0^{\circ}\)), respectively. HRIRs are transformed to HRTFs by applying the fast Fourier transform, with the resulting spectra limited to 0~Hz–20~kHz.  

In the azimuth plane, individual and non-individual measured HRTFs show rich direction-dependent spectral structure, including multiple pinna-related peaks and notches that shift systematically with azimuth. The individual 3D synthetic HRTFs appear to have slightly less pronounced direction-dependent spectral structure but seems to capture the individual pinna-related peaks and notches. The corresponding PR synthetic HRTFs exhibit smoother spectra with reduced dynamic range and fewer pronounced notches, especially above \(\approx 6\)~kHz. 

In the elevation plane, individual and non-individual measured HRTFs display moving high-frequency notches that encode elevation, whereas the synthetic representations appear smoothed with attenuated (3D) close to non-existent (PR) fine structure. These examples indicate that PR synthetic HRTFs capture directional trends but substantially reduced pinna-related peaks and notches.

\subsection{Average Metrics}

\begin{figure}[htbp]
    \centering
    \includegraphics[width=\columnwidth]{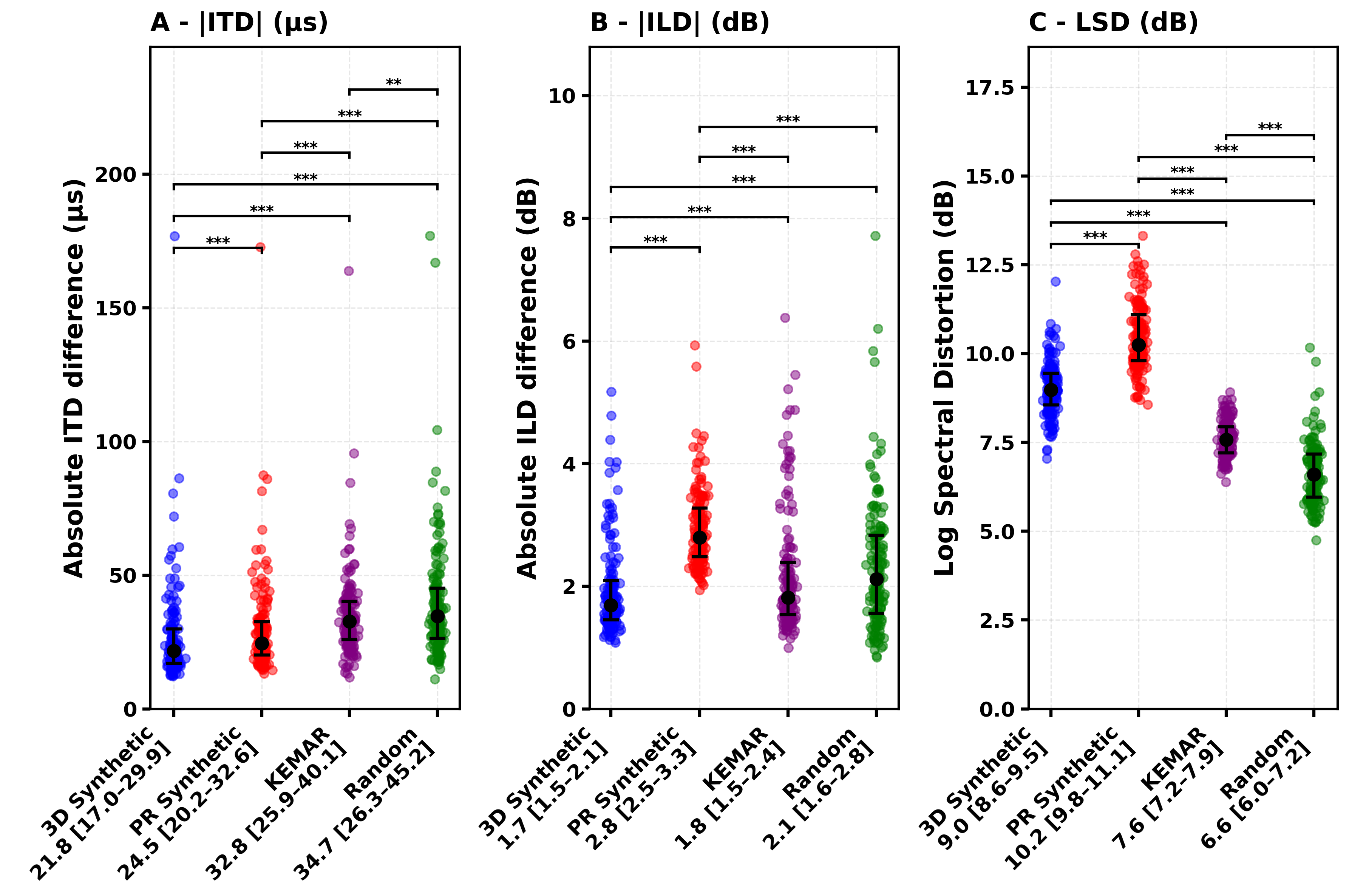}
    \caption{Average metrics across 150 subjects showing median [25th-75th percentile] with significance bars. A. Absolute ITD difference ($\mu$s), B. Absolute ILD difference (dB), C. Log-Spectral Distortion (dB). (* \(p < 0.05\), ** \(p < 0.01\), *** \(p < 0.001\)).}
    \label{fig:AverageMetrics}
\end{figure}

To quantify spectral differences observed in Figs.~\ref{fig:4Azimuth} and \ref{fig:4Elevation} and to evaluate the localisation cues differences across conditions, Figure \ref{fig:AverageMetrics} summarises subject-averaged metrics across 150 subjects for four conditions (3D synthetic, PR synthetic, KEMAR and Random) against measured HRTFs.

The statistical test revealed significant differences amongst conditions for all three metrics (\(p < 0.001\)). ITD differences remain low for all conditions (medians < 35 $\mu$s), with 3D synthetic HRTFs achieving the lowest ITD difference (21.8 $\mu$s), followed by the PR synthetic (24.5 $\mu$s), KEMAR (32.8 $\mu$s), and Random (34.7 $\mu$s) HRTFs. This indicates that individual head geometry (3D and PR synthetic) provides measurable benefits for ITD preservation over non-individual conditions.

3D synthetic HRTFs achieve lower ILD differences (1.7 dB) than all other conditions (\(p < 0.001\)). PR synthetic HRTFs exhibit elevated ILD differences (2.8 dB) compared to 3D synthetic (\(p < 0.001\)), KEMAR (1.8 dB, \(p < 0.001\)), and Random (2.1 dB, \(p < 0.001\)). KEMAR and Random conditions show no difference (\(p = 0.179\)).

LSD metrics reveal the most pronounced condition differences. Random HRTFs achieve the lowest spectral distortion (6.6 dB), followed by KEMAR (7.6 dB), 3D synthetic (9.0 dB), and PR synthetic (10.2 dB) (\(p < 0.001\)). Critically, PR synthetic HRTFs exhibit higher spectral distortion than all other conditions.

These aggregate metrics demonstrate that whilst PR synthetic HRTFs preserve ITD cues, they introduce substantial ILD and spectral errors that are even higher than those of non-individual HRTFs. 

\subsection{Spatial Distribution of Localisation Cues}

\begin{figure*}[htbp]
\centering
\includegraphics[width=0.95\textwidth]{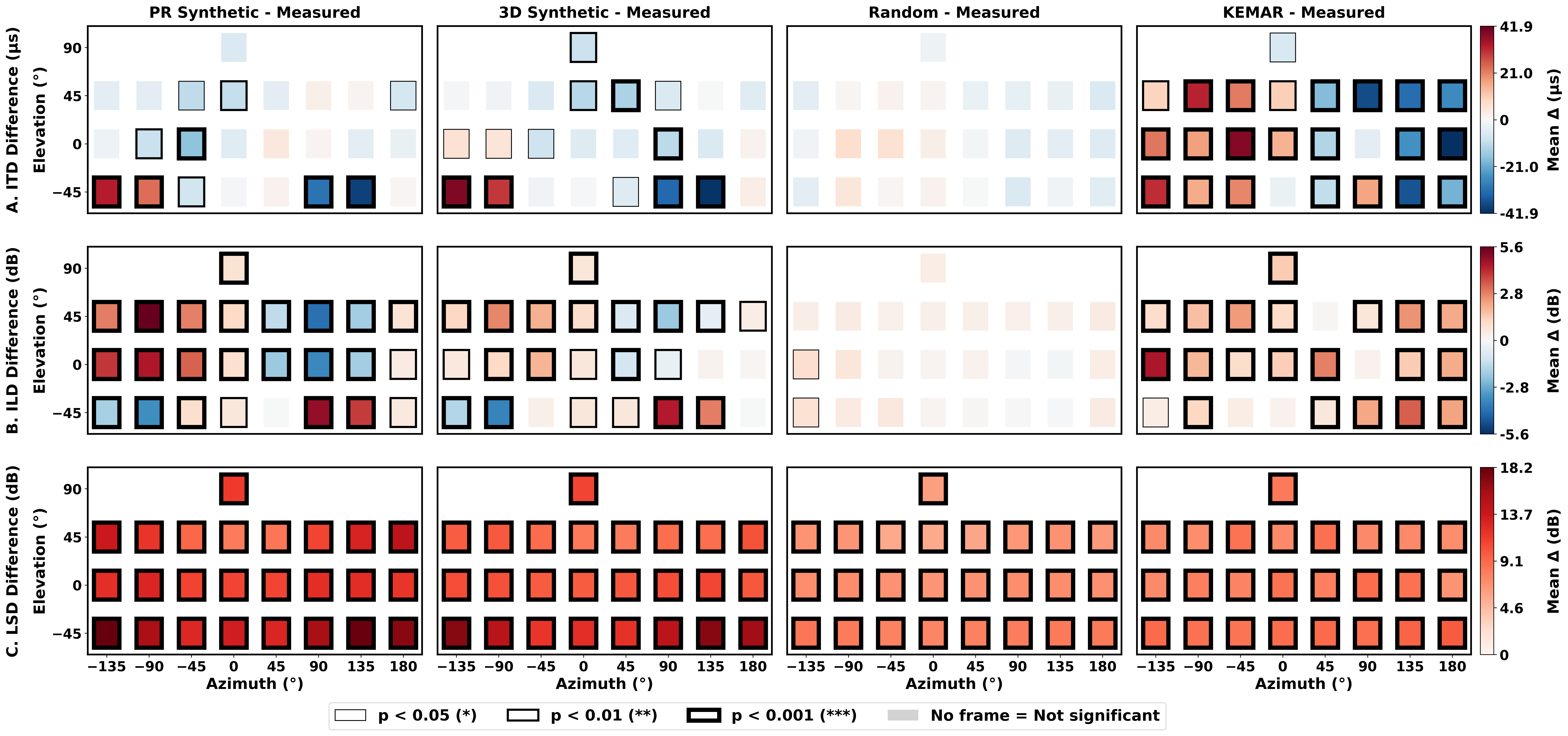}
\caption{Spatial distribution of ITD (top row), ILD (middle row), and LSD (bottom row) signed mean differences (condition minus measured) for four HRTF comparisons versus measured HRTFs, averaged over 150 subjects. The diverging colour map intensity represents signed mean error magnitude (Blue: Negative; Red: Positive). Black frames indicate positions where the group-mean difference is significantly different from zero after FDR correction across positions (thin: p < 0.05; medium: p < 0.01; thick: p < 0.001; no frame: not significant).}
\label{fig:SpatialMetrics}
\end{figure*}

The direction-dependence of HRTFs motivates the analysis of where these errors occur spatially and which locations show significant differences. The spatial distribution of ITD, ILD, and LSD differences across four comparisons (PR synthetic, 3D synthetic, Random, and KEMAR versus Measured HRTFs; condition minus measured), averaged over 150 subjects, is presented in Fig.~\ref{fig:SpatialMetrics}. The heatmaps show the signed mean difference ($\mu$s for ITD, dB for ILD and LSD), whilst boxes outlined in black indicate spatial locations where differences are significant or not with increasing thickness.

For PR synthetic HRTFs, ITD differences remain close to zero around frontal directions but show systematic biases at lateral azimuths and negative elevation positions, where PR synthetic HRTFs tend to underestimate ITDs on the contralateral side and overestimate them on the ipsilateral side. The 3D synthetic comparison shows a similar pattern compared to PR synthetic HRTFs. Random HRTFs exhibit no systematic ITD biases across all locations. A complementary unsigned analysis revealed that the Random condition yields larger, more widespread and less localised unsigned ITD differences than the other conditions. Across subjects, these ITD differences in the Random condition are not consistently biased in one direction relative to the measured HRTF so there is no consistent group-level bias in Fig.~\ref{fig:SpatialMetrics}. KEMAR HRTFs show systematic ITD differences, largely underestimating on the contralateral side and overestimating on the ipsilateral side. 

Regarding the ILD differences, PR synthetic HRTFs show a more widespread pattern of significance and higher signed magnitudes at lateral positions. The 3D synthetic comparison shows a similar pattern. Random HRTFs exhibit few ILD biases at the -135° lateral back location, and across subjects these ILD differences are not consistently biased in one direction. KEMAR HRTFs show positive ILD biases across the majority of positions, meaning KEMAR ILDs are systematically larger than the individual measured ILDs.

For LSD, all conditions showed significant deviations from the measured HRTFs at all positions (FDR-corrected (p < 0.001)). The magnitude and spatial pattern of LSD are more informative than the significance coding in this case. The PR synthetic comparison shows a similar spatial pattern of LSD differences compared to the 3D synthetic HRTFs with higher LSD at negative elevation positions. Random HRTFs show LSD errors that are lower across all locations compared to both synthetic conditions and KEMAR, with LSD errors for KEMAR evenly spatially distributed across locations similarly to Random HRTFs. 

These spatial patterns reveal that whilst synthetic HRTFs preserve ITD cues reasonably well as indicated in Fig.~\ref{fig:AverageMetrics}, they introduce spectral errors (LSD) that are spatially distributed and highest at negative elevation positions. Synthetic HRTFs show localised ITD and ILD biases at lateral azimuths and negative elevation positions. KEMAR HRTFs, in contrast, introduce a systematic positive ILD bias across nearly all spatial positions, indicating that the mannequin's head geometry produces consistently larger interaural level differences than individual subjects. KEMAR imposes a global unidirectional scaling of interaural time differences for 150 subjects. Random HRTFs, while exhibiting LSD errors at all positions, do not produce consistent group-level ITD or ILD biases in one direction but show higher and more spatially distributed unsigned ITD differences. The unsigned ILD distribution reveals more spatially widespread differences for Random HRTFs compared to other conditions. 

\subsection{Frequency-Dependent Spectral Distortion}

\begin{figure*}[htbp]
\centering
\includegraphics[width=0.8\textwidth]{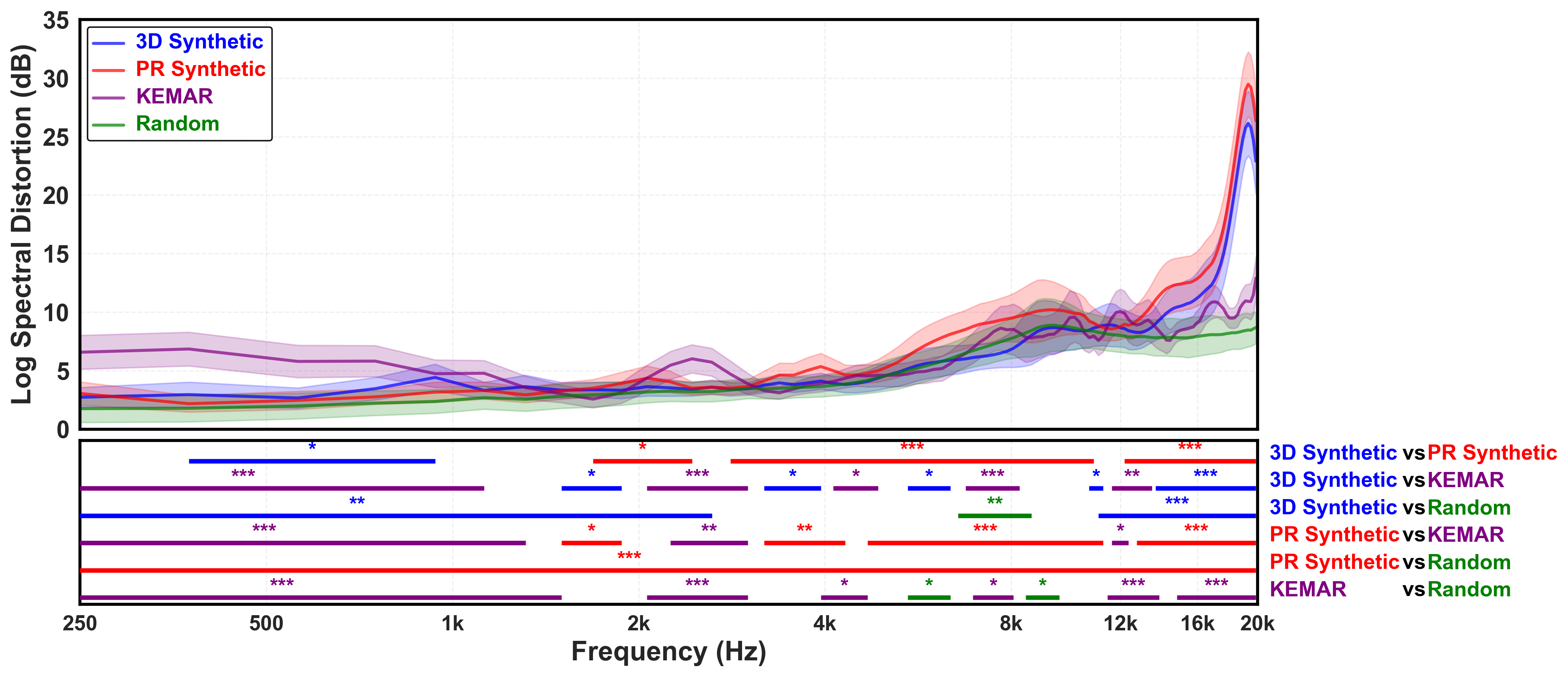}
\caption{Frequency-dependent log-spectral distortion across 150 subjects. Top: Mean LSD curves with standard deviation bands for four HRTF conditions compared to measured HRTFs. Bottom: Cluster permutation statistics showing frequency regions where each row corresponds to one pairwise comparison. For a given bar, coloured segments mark the frequency clusters where that condition has a significantly higher LSD than its counterpart. Asterisks indicate significance levels (* \(p < 0.05\), ** \(p < 0.01\), *** \(p < 0.001\)).}
\label{fig:LSD}
\end{figure*}

To identify which frequency regions show significant differences between HRTF conditions and the measured HRTF, LSD values were averaged across ears, directions, and subjects (N = 150) for the four HRTF conditions (3D synthetic, PR synthetic, KEMAR, and Random; see Fig.~\ref{fig:LSD}. The upper panel shows the average LSD curves for each condition, whilst the lower panel displays cluster-based permutation statistics identifying significantly difference frequency clusters (\(p < 0.05\)). Horizontal bars indicate frequency clusters corresponding to the condition exhibiting the higher LSD for each of the six pairwise comparisons. 

Across all conditions, the LSD remains relatively low at low frequencies (< 500 Hz) except for the KEMAR HRTF, which shows elevated distortion compared to all other conditions from 250 Hz onwards (purple bar, \(p < 0.001\)). Above 2 kHz, the LSD increases progressively for all conditions, with the steepest rise occurring above 12 kHz for the synthetic conditions.

3D synthetic HRTFs show lower distortion than PR synthetic HRTFs across multiple frequency clusters: \(\approx 2\)~kHz (\(p < 0.05\)), 3–10 kHz (\(p < 0.001\)), and 12–20 kHz (\(p < 0.001\)), indicating that higher mesh resolution provides consistent benefits across a broad frequency range. PR synthetic HRTFs exhibit higher distortion than Random HRTFs across the entire spectrum. Whilst both 3D and PR synthetic HRTFs follow Random HRTFs closely in mid-frequencies (as observed in the overlapping curves), they diverge above \(\approx 11\)~kHz, with PR synthetic showing the most severe degradation.

Overall, frequency-dependent spectral distortion analysis reveals that PR synthetic HRTFs show higher distortion than 3D synthetic HRTFs especially after 1 kHz.  Beyond 12 kHz to 20 kHz both synthetic conditions deviate much more than the KEMAR and Random conditions.

\subsection{Auditory Models}

\subsubsection{Baumgartner2014 Model}

\begin{figure*}[htbp]
    \centering
    \includegraphics[width=0.95\textwidth]{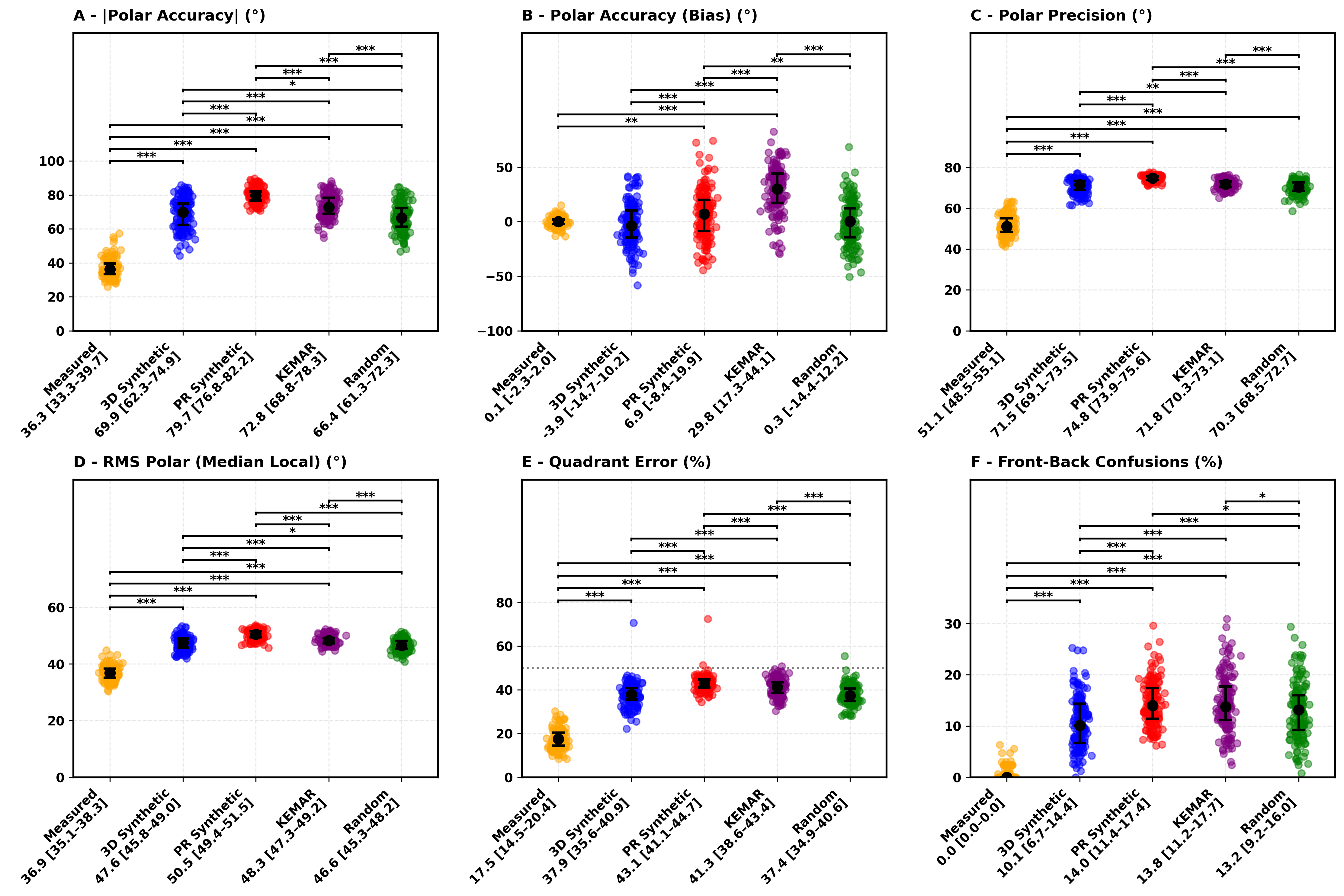}
    \caption{Baumgartner2014 model predictions across 150 subjects showing median [25th-75th percentile] with significance bars. Metrics: A. |Polar Accuracy| (°), B. Polar Accuracy (Bias) (°), C. Polar Precision (°), D. RMS Polar (Median Local) (°), E. Quadrant Error (\%), F. Front-Back Confusions (\%). (* \(p < 0.05\), ** \(p < 0.01\), *** \(p < 0.001\)).}
    \label{fig:B2014}
\end{figure*}

To assess whether any observed numerical differences translate into predicted localisation performance differences, model-based assessment was performed on the different HRTF sets. The Baumgartner2014 model predicts localisation performance across six metrics showing median [25th-75th percentile] with significance bars (Fig.~\ref{fig:B2014}). The measured HRTF baseline (template = target) yields the most accurate predictions: |polar accuracy| = 36.3°, quadrant error = 17.5\%, front-back confusions = 0.0\%.

For |polar accuracy|, 3D synthetic (69.9°) and Random (66.4°) HRTFs demonstrate better predicted performance than KEMAR (72.8°, \(p < 0.001\)) and PR synthetic (79.7°, \(p < 0.001\)). The 3D synthetic and Random conditions show small difference in predicted polar accuracy (\(p = 0.029\)). PR synthetic predictions exhibit the poorest polar accuracy, performing worse than all other non-individual conditions (\(p < 0.001\)).

Regarding predicted confusion rates, 3D synthetic HRTFs yield lower front-back confusions (10.1\%) compared to PR synthetic (14.0\%, \(p < 0.001\)), KEMAR (13.8\%, \(p < 0.001\)), and Random (13.2\%, \(p < 0.001\)). For quadrant errors, 3D synthetic (37.9\%) and Random (37.4\%) show statistically equivalent predicted performance (\(p = 0.752\)), both outperforming PR synthetic (43.1\%) and KEMAR (41.3\%) with \(p < 0.001\).

\subsubsection{Barumerli2023 Model}

\begin{figure*}[htbp]
    \centering
    \includegraphics[width=0.95\textwidth]{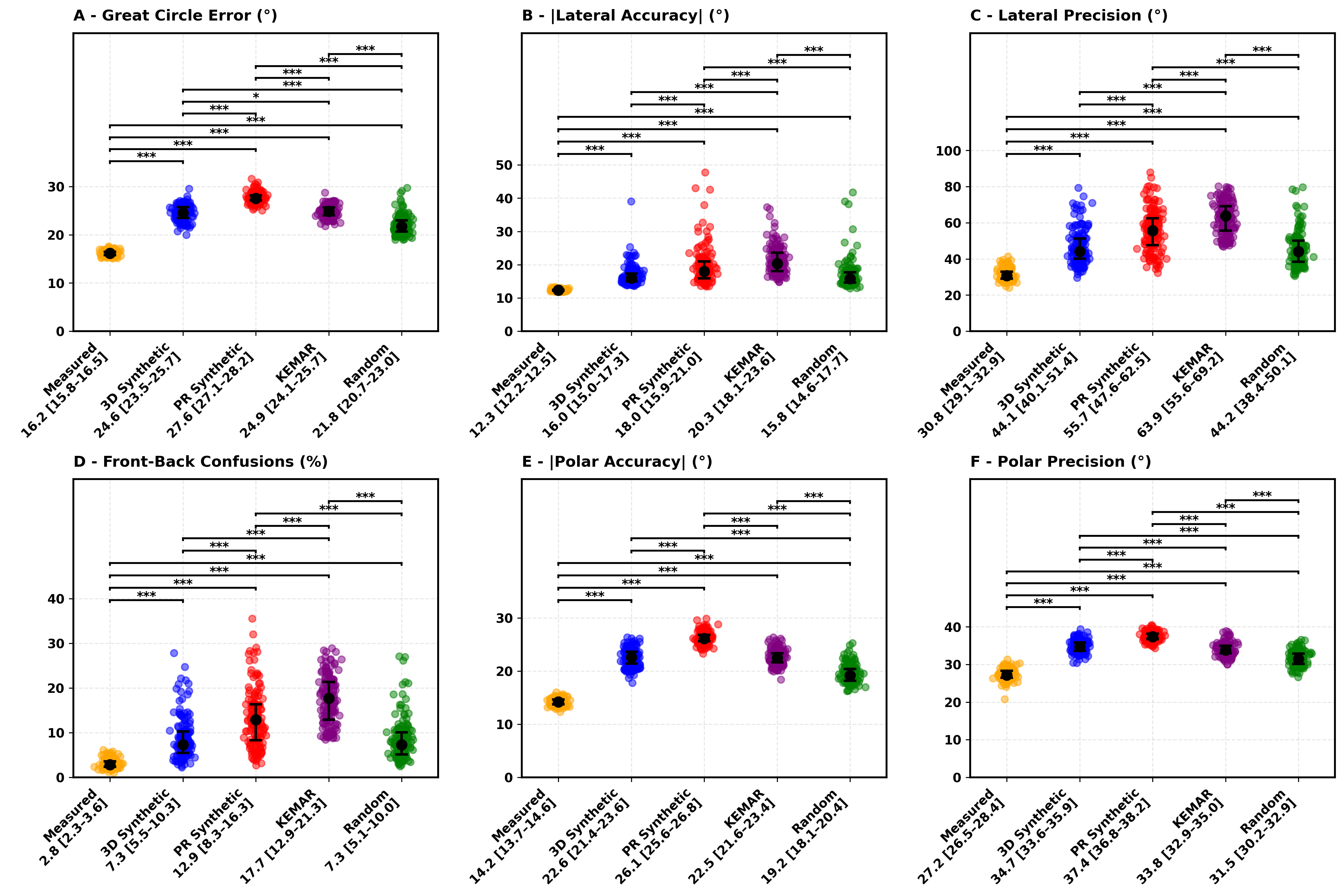}
    \caption{Barumerli2023 model predictions across 150 subjects showing median [25th-75th percentile] with significance bars. Metrics: A. Great Circle Error (°), B. |Lateral Accuracy| (°), C. Lateral Precision (°), D. Front-Back Confusions (\%), E. |Polar Accuracy| (°), F. Polar Precision (°). (* \(p < 0.05\), ** \(p < 0.01\), *** \(p < 0.001\)).}
    \label{fig:B2023}
\end{figure*}

The Barumerli2023 model predictions across six metrics are presented in Fig.~\ref{fig:B2023} showing median [25th-75th percentile] with significance bars. The measured HRTF baseline demonstrates the lowest predicted errors: great circle error = 16.2°, |polar accuracy| = 14.2°, front-back confusions = 2.8\%.

For predicted |polar accuracy|, Random HRTFs (19.2°) outperform 3D synthetic (22.6°), PR synthetic (26.1°) and KEMAR (22.5°) with \(p < 0.001\). Notably, 3D synthetic and KEMAR predictions show no difference (\(p = 0.671\)). PR synthetic predictions exhibit higher |polar accuracy| errors (26.1°) compared to all other conditions (\(p < 0.001\)). The Barumerli2023 model predicts systematically lower absolute polar errors than Baumgartner2014 across all HRTF conditions.

For front-back confusions, 3D synthetic and Random HRTFs demonstrate statistically equivalent predicted performance. These conditions yield lower predicted front-back confusion rates than PR synthetic and KEMAR (\(p < 0.001\)). Specifically, predicted front-back confusions for 3D synthetic (7.3\%) and Random (7.3\%) are approximately half those predicted for KEMAR (17.7\%) and lower than PR synthetic (12.9\%, \(p < 0.001\)). 

\begin{figure}[htbp]
    \centering
    \includegraphics[width=0.9\columnwidth]{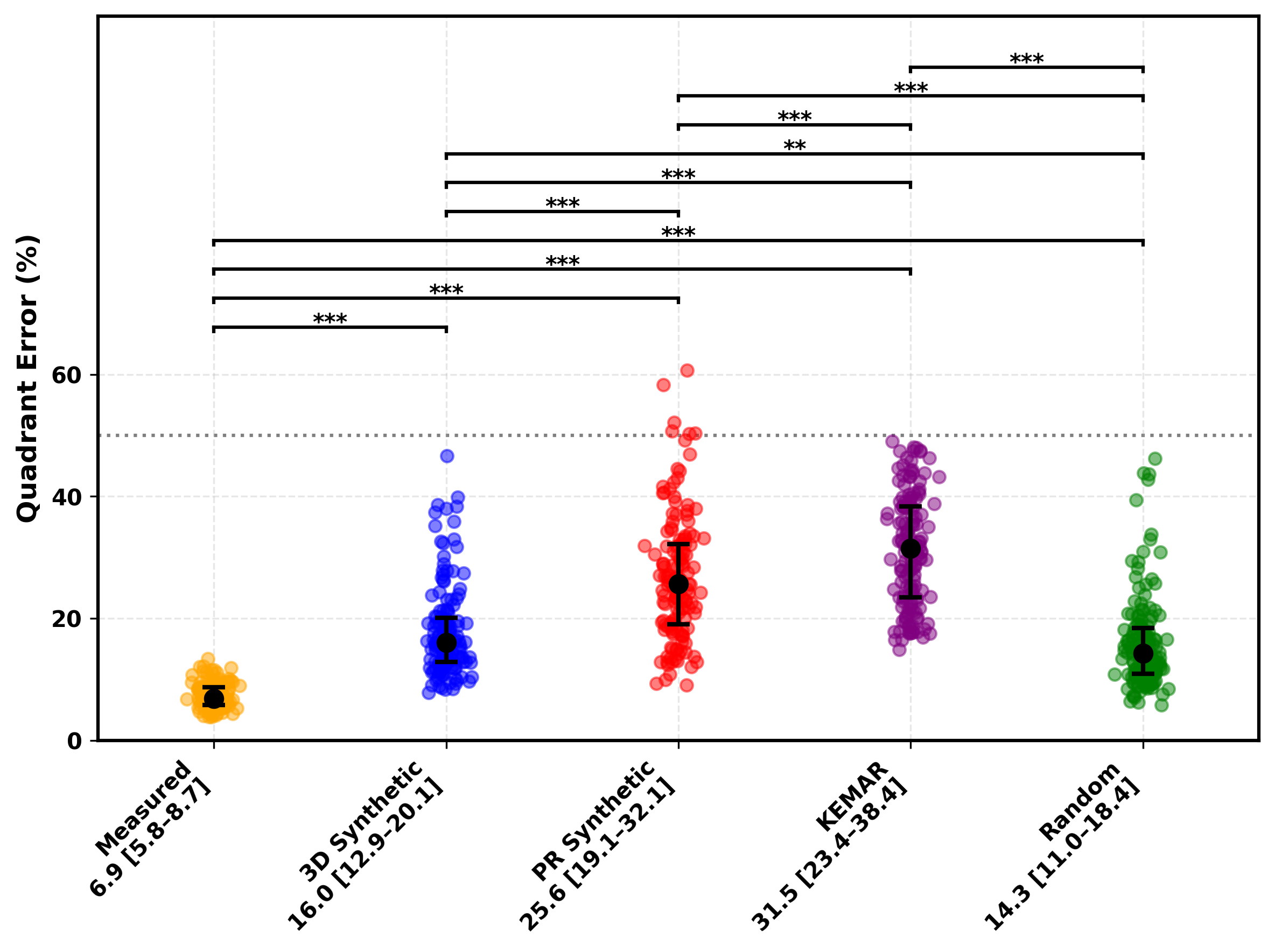}
    \caption{Barumerli2023 model predicted quadrant error rate across 150 subjects showing median [25th-75th percentile] with significance bars (* \(p < 0.05\), ** \(p < 0.01\), *** \(p < 0.001\)).}
    \label{fig:B2023Querr}
\end{figure}

Predicted quadrant error rate from the Barumerli2023 model are illustrated in Fig.~\ref{fig:B2023Querr}. The measured HRTF baseline yields the lowest predicted quadrant errors (6.9\%), establishing reference performance. Amongst the other conditions, Random HRTFs demonstrate the best predicted performance (14.3\%), followed by 3D synthetic (16.0\%), PR synthetic (25.6\%) and KEMAR (31.5\%). Statistical analysis reveals that 3D synthetic HRTFs exhibit lower predicted quadrant errors than PR synthetic (\(p < 0.001\)) and KEMAR (\(p < 0.001\)), but higher errors than Random HRTFs (\(p = 0.005\)). PR synthetic predictions are worse than all other conditions (\(p < 0.001\)) except KEMAR that exhibits the highest predicted quadrant error rate.

Overall, both models agree on the ranking of conditions: measured HRTFs results in best predictions, followed by 3D synthetic and Random, with PR synthetic and KEMAR revealing worse predictions. PR synthetic HRTFs are consistently predicted to perform worst or second-worst, especially on polar accuracy, quadrant error, and front-back confusions. These predictions are, however, based on auditory models that are calibrated for measured HRTFs and have not been validated specifically for synthetic HRTFs. Assessment of the localisation cues provided by these HRTFs for human listeners therefore still requires direct behavioural testing in a sound localisation task.
 
\subsection{Localisation Test}

\begin{figure*}[htbp]
    \centering
    \includegraphics[width=0.85\textwidth]{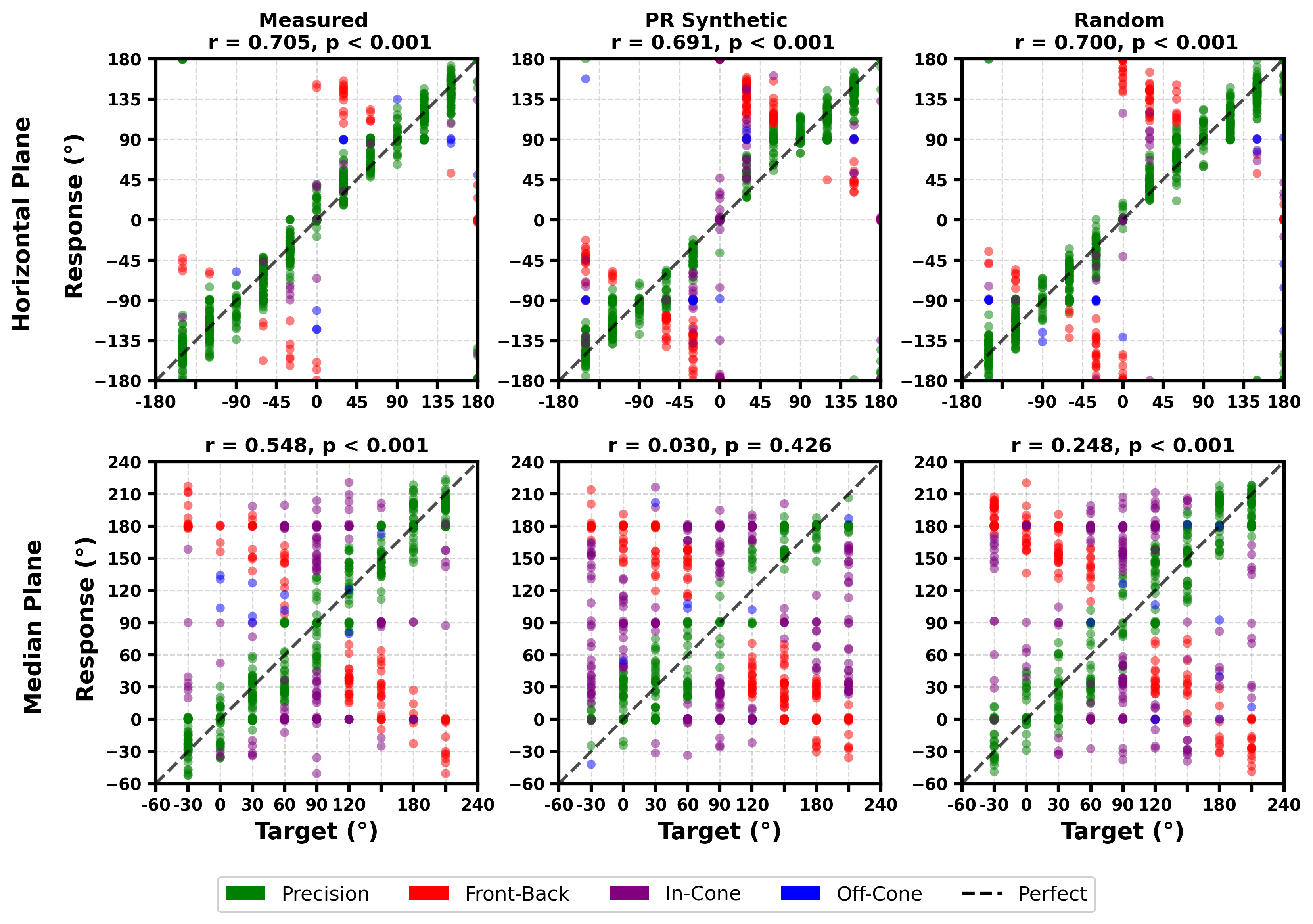}
    \caption{Raw localisation responses for 27 participants in horizontal (top row) and median (bottom row) planes. Each point represents a single trial coloured by confusion type (green = precision, red = front–back, purple = in-cone, blue = off-cone). Dashed line indicates perfect localisation. Pearson correlation coefficients (r) with significance levels are shown for each condition.}
    \label{fig:LocaRaw}
\end{figure*}

To assess listeners' localisation performance with synthetic HRTFs, a behavioural listening test was conducted with 27 normal-hearing participants employing acoustically measured, PR synthetic, and randomly selected HRTFs. Localisation responses in the horizontal and median planes for the three HRTF conditions are shown in Fig.~\ref{fig:LocaRaw}, with Pearson correlation coefficients quantifying the linear relationship between target and response positions.

Correlation analysis reveals distinct performance patterns across conditions and spatial planes. In the horizontal plane, all HRTF conditions show strong correlations (\(p < 0.001\)): Measured (\(r = 0.705\)), PR Synthetic (\(r = 0.691\)) and Random (\(r = 0.700\)). These values indicate a strong linear trend between target and response in the horizontal plane for all three conditions. The extent to which errors reflect random variability versus systematic biases (e.g. front–back confusions) can be seen in the raw data in Fig.~\ref{fig:LocaRaw}.

In contrast, the median plane shows markedly divergent performance. Measured HRTFs exhibit moderate correlation (\(r = 0.548, p < 0.001\)), indicating preserved elevation perception despite increased variability. Random HRTFs show weak but significant correlation (\(r = 0.248, p < 0.001\)), suggesting substantial elevation confusion. Critically, PR Synthetic HRTFs show no correlation (\(r = 0.030, p = 0.426\)), indicating near-random elevation responses widely distributed across incorrect quadrants and frequent front–back confusions.

\begin{figure*}[htbp]
    \centering
    \includegraphics[width=0.95\textwidth]{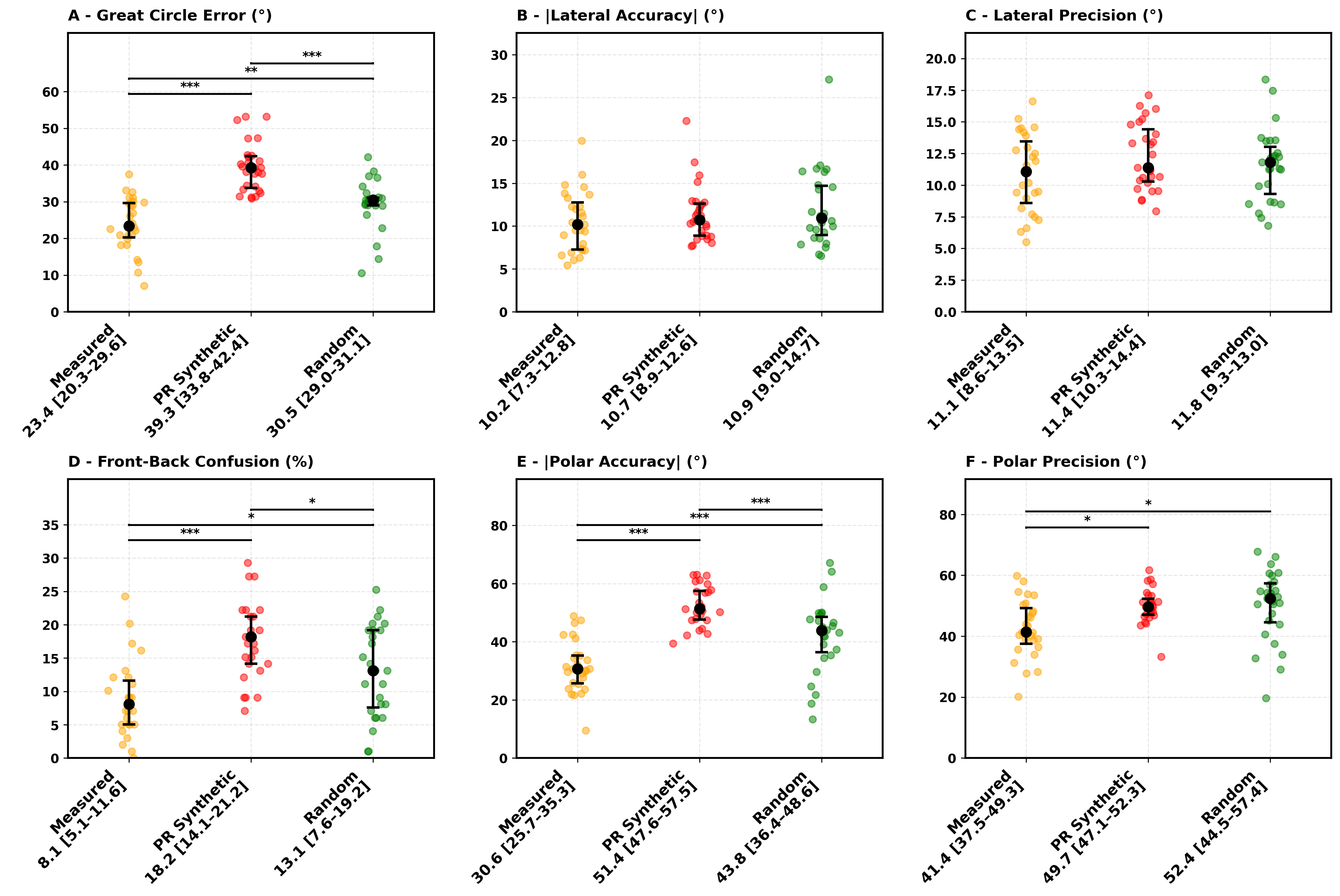}
    \caption{Localisation performance metrics for 27 participants showing median [25th-75th percentile] with significance bars for Measured, PR synthetic, and Random HRTFs. A. Great-circle error (°), B. Absolute lateral accuracy (°), C. Lateral precision (°), D. Front-back confusions (\%), E. Absolute polar accuracy (°), F. Polar precision (°). (* \(p < 0.05\), ** \(p < 0.01\), *** \(p < 0.001\)).}
    \label{fig:LocaPerf}
\end{figure*}

Aggregate localisation metrics in Fig.~\ref{fig:LocaPerf} quantify performance differences observed in the correlation analysis across six metrics showing median [25th-75th percentile] with significance bars. Measured individual HRTFs yield lower great-circle error (23.4°) than both PR synthetic (39.3°, \(p < 0.001\)) and Random HRTFs (30.5°, \(p < 0.01\)), as shown in Fig.~\ref{fig:LocaPerf}A. PR synthetic HRTFs produced higher errors than Random HRTFs (\(p < 0.001\)). 

Azimuthal localisation remained relatively robust across HRTF types, consistent with the horizontal plane correlations. Absolute lateral accuracy showed no significant differences amongst measured (10.2°), PR synthetic (10.7°), and Random (10.9°) conditions (Fig.~\ref{fig:LocaPerf}B). Similarly, lateral precision exhibited no differences between HRTF conditions (Fig.~\ref{fig:LocaPerf}C). In contrast, elevation perception showed marked degradation, corroborating the weak median plane correlations. Measured HRTFs yielded lower absolute polar accuracy (30.6°) than both PR synthetic (51.4°, \(p < 0.001\)) and Random HRTFs (43.8°, \(p < 0.001\)), as illustrated in Fig.~\ref{fig:LocaPerf}E. PR synthetic HRTFs exhibited higher polar errors than Random HRTFs (\(p < 0.001\)). Polar precision (Fig.~\ref{fig:LocaPerf}F) revealed that PR synthetic and Random HRTFs showed higher variability compared to measured HRTFs (\(p < 0.05\)).

Front–back confusion rates (Fig.~\ref{fig:LocaPerf}D) were lowest for measured HRTFs (8.1\%) and higher for both PR synthetic (18.2\%, \(p < 0.001\)) and Random (13.1\%, \(p < 0.05\)) conditions, with significant difference between the latter two (\(p < 0.05\)). These elevated confusion rates corroborate the weak median plane correlations observed for PR synthetic and Random HRTFs.

\begin{figure}[htbp]
    \centering
    \includegraphics[width=0.9\columnwidth]{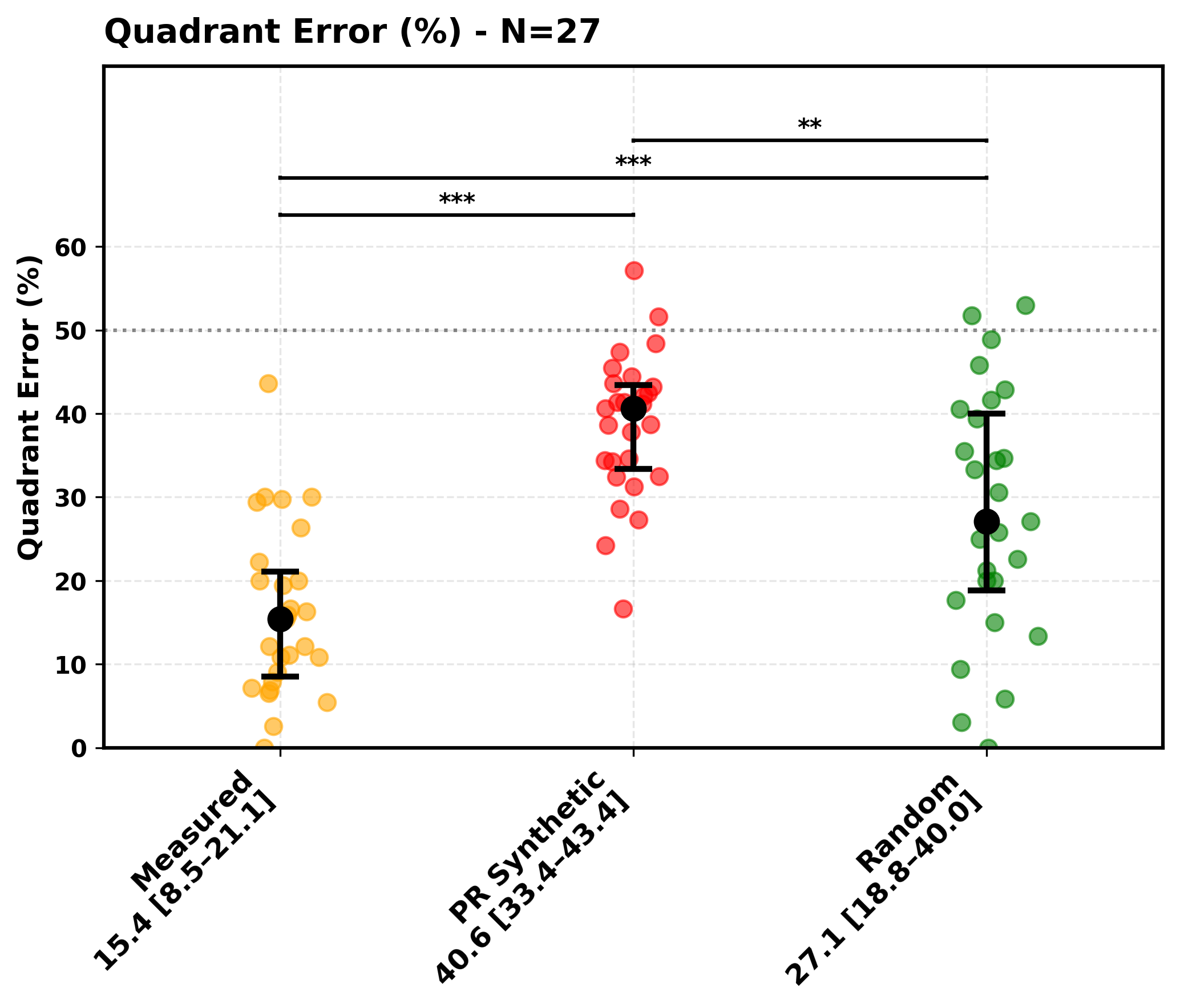}
    \caption{Quadrant error metric for 27 participants showing median [25th-75th percentile] with significance bars for Measured, PR synthetic, and Random HRTFs. (* \(p < 0.05\), ** \(p < 0.01\), *** \(p < 0.001\)).}
    \label{fig:ConfusionMetrics}
\end{figure}

Figure~\ref{fig:ConfusionMetrics} shows the quadrant error (QE) confusion metric results with values reported as median [25th–75th percentile] of participant medians. The measured HRTF condition achieved lower rates (15.4\%) to PR synthetic (40.6\%) and to Random (27.1\%) (\(p < 0.001\)) with significant difference between the latter two (\(p < 0.01\)).

The localisation test shows that elevation-related localisation (polar accuracy, front–back and quadrant confusions) is strongly degraded for PR synthetic and, to a lesser extent, for Random HRTFs, while azimuth remains robust across conditions, and that PR synthetic HRTFs yield the poorest overall localisation performance.

\subsection{Spectral Similarity and Behavioural Performance}

In the localisation experiment, each participant was assigned a different randomly selected measured HRTF from the SONICOM dataset as the non-individual condition, so that the benchmark reflects a non-individual HRTF rather than a single mannequin. A potential concern with this approach is that spectral similarity between each participant's measured HRTF and their assigned random HRTF varies across participants (some may receive an HRTF spectrally close to their own, others one that is spectrally more different). If that similarity influenced performance, the Random condition would be an unfair or inconsistent benchmark. To test whether spectral distance predicts behavioural outcome, LSD between each participant's measured HRTF and their assigned random HRTF was correlated with the performance difference (Random minus Measured) for four behavioural metrics. For PR synthetic HRTFs, each participant received their own, so there is no between-participant variation in which PR HRTF is assigned. The present analysis therefore focuses on the Random condition, where assignment varies. LSD values ranged from 5.26 to 8.78~dB (median: 7.14 \(\pm\) 0.77~dB) across the 27 participants, indicating relatively modest inter-participant variation in spectral distance to the assigned random HRTF.

\begin{figure*}[htbp]
    \centering
    \includegraphics[width=\textwidth]{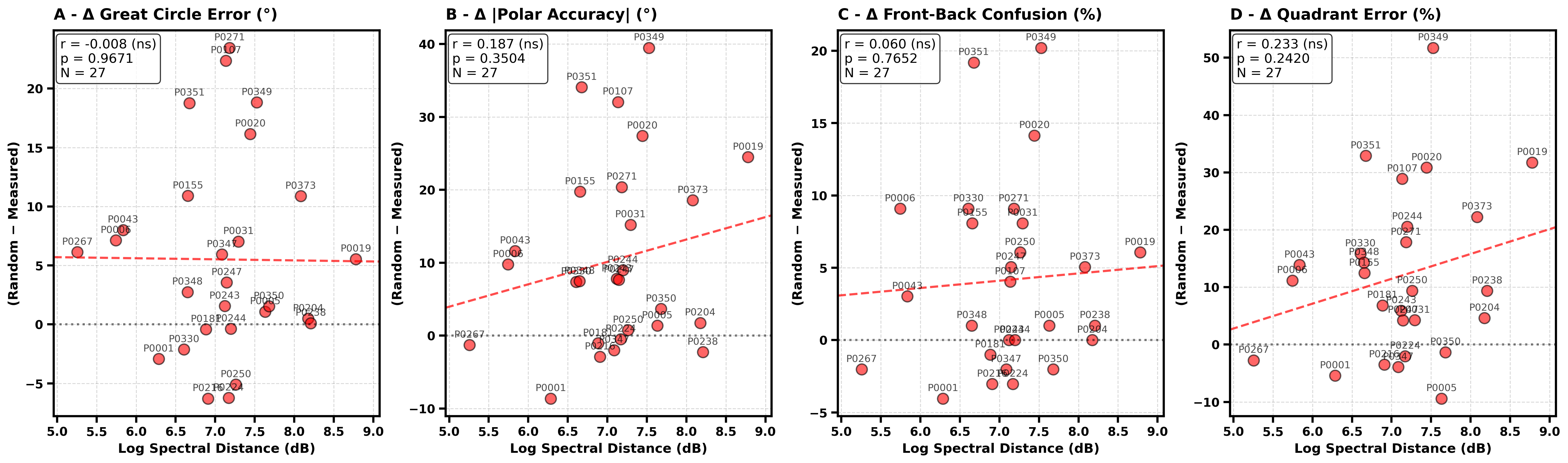}
    \caption{Correlation between log-spectral distortion (1–16~kHz) and behavioural performance differences (Random minus Measured) for 27 participants.}
    \label{fig:LSDCorrelation}
\end{figure*}

Figure~\ref{fig:LSDCorrelation} shows the relationship between LSD and performance differences for each metric. Each red point represents one participant. Dashed red lines indicate linear regression fits. No significant (ns) correlations were observed: great-circle error (\(r = -0.008\), \(p = 0.9671\)), absolute polar accuracy (\(r = 0.187\), \(p = 0.3504\)), front-back confusion rate (\(r = 0.060\), \(p = 0.7652\)), and quadrant error rate (\(r = 0.233\), \(p = 0.2420\)). The absolute lateral accuracy is not shown as no significant correlation was obtained (\(r = -0.064\), \(p = 0.7498\)). All correlation coefficients were near zero (\(|r| < 0.25\)), indicating no linear relationship between spectral distance and behavioural performance differences. Substantial individual variability was observed, with some participants performing better with the random HRTF despite high LSD values, whilst others performed worse despite low LSD values. These findings suggest that simple spectral similarity alone does not predict HRTF compatibility for localisation tasks and support the use of a different randomly selected non-individual HRTF per participant as a valid benchmark for the Random condition.

\section{Discussion}

\subsection{RQ1: Numerical Fidelity of PR Synthetic HRTFs}

Statistical analysis across 150 subjects demonstrates that PR synthetic HRTFs preserve interaural timing cues but exhibit degraded ILDs and spectral fidelity. Aggregate metrics (Fig.~\ref{fig:AverageMetrics}) demonstrate preserved ITD accuracy for PR synthetic and 3D synthetic HRTFs, both outperforming non-individual conditions. However, ILD errors for PR synthetic are higher compared to all other conditions. Critically, LSD analysis reveals PR synthetic HRTFs exhibit higher spectral distortion than all other conditions, with all pairwise comparisons reaching significance. LSD values for non-individual measured HRTFs (median 6.6~dB) are in line with values obtained in past research \cite{hogg_hrtf_2024, hu_hrtf_2024, hu_machine_2025} on the SONICOM dataset. The alignment with prior literature supports the method used to obtain these metric results and their validity. 

Spatial analysis (Fig.~\ref{fig:SpatialMetrics}) with permutation testing identifies ITD and ILD deviations concentrated at lateral azimuths (\(\pm\)90°, \(\pm\)135°) and negative elevation angles. The absence of a torso in the 3D meshes could potentially explain the negative elevation errors, as torso and shoulder geometry are known to contribute to elevation-dependent cues and to improve congruence with measured HRTFs in the vertical direction \cite{algazi_approximating_2002, kreuzer_fast_2009}. 

Frequency-dependent cluster permutation analysis (Fig.~\ref{fig:LSD}) indicates that PR synthetic HRTFs exhibit higher distortion than 3D synthetic HRTFs, particularly after 1 kHz. Beyond 12 kHz to 20 kHz, both synthetic conditions deviate more compared to the KEMAR and Random conditions. This high deviation from the synthetic conditions could be due to limitations in the BEM calculations, as the wavelength decreases and the mesh resolution becomes insufficient \cite{ziegelwanger_calculation_2013, kreuzer_fast_2009} and also the difference of magnitude between the measured and synthetic HRTFs at high frequencies due to the HRTF measurement microphone frequency response \cite{engelSONICOMHRTFDataset2023}. 

Spectral heatmaps (Figs.~\ref{fig:4Azimuth}, \ref{fig:4Elevation}) indicate that measured HRTFs exhibit direction-dependent fine structure with prominent pinna-related notches, whilst PR synthetic HRTFs display smoothed characteristics with attenuated high-frequency features. This spectral degradation, quantified through statistical analysis, directly reflects coarse pinna geometry in PR meshes and critically impairs monaural cues. 

In answer to RQ1, PR synthetic HRTFs do not preserve the key numerical characteristics of measured HRTFs to a similar extent as 3D synthetic HRTFs. The present work focuses on a single PR and mesh-processing pipeline and on a specific HRTF synthesis method (Mesh2HRTF) and measurement configuration. The results may therefore not generalise to alternative reconstruction algorithms, mesh-optimisation strategies, or numerical solvers. Further work to close this gap would need to improve capture of fine pinna detail (e.g.\ through higher-resolution reconstruction or targeted ear refinement), consider inclusion of torso geometry in the simulation, or combine PR with complementary individualisation steps. Whether the observed numerical degradations translate into predicted or actual perceptual impairments is addressed by the auditory modelling (RQ2) and behavioural (RQ3) assessments below.

\subsection{RQ2: Predicted Perceptual Impact}

Auditory model predictions across 150 subjects indicate that numerical degradation in PR synthetic HRTFs translates to substantial predicted localisation impairment. Both Baumgartner2014 and Barumerli2023 models predict poorest performance for PR synthetic HRTFs across all metrics. Specifically, Baumgartner2014 predicts that PR synthetic polar accuracy is worse than 3D synthetic , KEMAR and Random. Barumerli2023 predictions corroborate this pattern. Baumgartner2014 predicts PR synthetic front-back confusions  and higher quadrant errors than 3D synthetic and Random conditions. Barumerli2023 similarly predicts PR synthetic higher front-back confusions than 3D synthetic and Random. 

Model predictions should be interpreted within the context of the specific parameter configurations and HRTF sets employed here, as direct quantitative comparison with prior studies \cite{hogg_hrtf_2024,baumgartner_modeling_2014, barumerli_bayesian_2023, daugintis_classifying_2023} is precluded by differences in model parameters, template/target HRTF selections, and lateral angle sampling (Baumgartner2014). Both models are developed primarily for acoustically measured HRTFs and therefore may not fully capture the perceptual implications of the specific spectral characteristics introduced by synthetic HRTFs.

In the context of RQ2, this caveat applies to cross-study comparison of predicted values. Within the present approach, the convergent ranking of conditions (PR synthetic worst, then KEMAR, 3D synthetic and Random) across both models directly supports the conclusion that the observed numerical degradations in PR synthetic HRTFs do translate into predicted localisation impairment particularly for elevation-dependent and front–back disambiguation tasks. Whether these auditory model predictions translate to actual behavioural localisation deficits with human listeners required direct experimental validation.

\subsection{RQ3: Behavioural Localisation Performance}

Addressing the third research question, the virtual-reality localisation experiment with 27 participants confirms that PR synthetic HRTFs do not yet match the perceptual performance afforded by measured HRTFs. Pearson correlation analysis (Fig.~\ref{fig:LocaRaw}) demonstrates preserved azimuthal localisation across conditions, but divergent elevation performance. The behavioural data align with the numerical and model-based findings. When rendered with their PR synthetic HRTFs, participants yield substantially larger great-circle and elevation errors and higher front-back and quadrant error rates than with measured or randomly assigned non-individual HRTFs, whilst azimuthal performance remains comparable across conditions. Thus the numerical degradations identified in RQ1 and the predicted impairments from RQ2 are confirmed in localisation behaviour. The sound localisation test results, values and patterns are in line with previous studies comparing individual and non-individual HRTFs \cite{meyer_generalization_2025,daugintis_initial_2023,middlebrooks_virtual_1999,majdak_acoustic_2014, majdak_two-dimensional_2011}. 

Critically, azimuthal metrics show no differences amongst conditions, confirming that ITD preservation maintains lateral localisation whilst spectral degradation selectively impairs elevation perception, in line with prior work \cite{middlebrooks_sound_1991, langendijk_contribution_2002, baumgartner_modeling_2014}, similarly identified in the numerical and model-based assessments. 

The correlation analysis (Fig.~\ref{fig:LSDCorrelation}) revealed no significant relationship between spectral similarity and performance differences, suggesting that individual variability in HRTF compatibility reflects factors beyond simple spectral distance. It follows that LSD, whilst useful for characterising group-level spectral differences between HRTF sets, should not be over-interpreted as a predictor of localisation performance for PR synthetic or other non-individual HRTFs. Monaural cues are difficult to reduce to a single global metric, they are direction- and frequency-dependent \cite{langendijk_contribution_2002, baumgartner_modeling_2014}, and listeners weight specific spectral regions \cite{llado_spectral_2025}. A spatially and spectrally averaged LSD over all directions, ears and frequencies therefore does not correlate with localisation performance. Future work could test whether position- and frequency-specific spectral deviations correlate with specific localisation errors, which would better align metric and perception.

\subsection{Outlook}

The findings above establish a practical baseline and a comprehensive evaluation framework that combines numerical metrics, auditory models, and behavioural data. Concrete directions for improvement are now well defined. On the geometric side, improvements in image capture protocols, reconstruction algorithms, and curvature-adaptive re-meshing could enhance ear detail whilst maintaining reasonable mesh size. The inclusion of torso geometry in future simulations may further reduce elevation-related deviations. 

On the acoustic side, hybrid approaches that combine PR synthetic HRTFs with machine-learning models to restore high-frequency spectral detail from ear images may help bridge the gap to measured HRTFs. Relevant methods include image-based high-frequency monaural cues prediction \cite{zhao_magnitude_2022, lee_personalized_2018}, DNN-based individualisation from pinna-related transfer functions \cite{prtfnet_2024}, and neural network prediction of HRTFs from 3D head meshes \cite{zhao_hrtf_3dmesh_2024}. Together, these directions offer a clear path toward narrowing the remaining gap in spectral and perceptual fidelity for accessible individual HRTF synthesis.

\section{Conclusion}

This study evaluated whether photogrammetry reconstruction can provide a practically useful basis for individual HRTF synthesis by addressing three research questions concerning numerical fidelity, predicted perceptual impact and behavioural localisation performance. Using photogrammetry data from the SONICOM dataset, individual synthetic HRTFs were generated for 150 subjects and evaluated against measured HRTFs, 3D synthetic HRTFs, KEMAR, and randomly selected HRTFs through numerical metrics, auditory models, and a virtual-reality sound localisation experiment.

Regarding numerical fidelity (RQ1), PR synthetic HRTFs preserve overall interaural timing cues but exhibit increased interaural level differences and spectral distortion relative to high resolution 3D scan synthetic HRTFs. Regarding predicted perceptual impact (RQ2), auditory-model predictions indicate that these differences translate into substantially higher elevation errors, front-back confusions and quadrant error rates compared to non-individual and individual measured HRTFs. Regarding behavioural performance (RQ3), the localisation experiment confirms that PR synthetic HRTFs yield worse elevation accuracy and confusions rates than measured HRTFs and randomly selected HRTFs.

Nevertheless, this work establishes a practical, end-to-end baseline pipeline for accessible individual HRTF synthesis and a comprehensive evaluation framework that combines numerical metrics, auditory models, and behavioural data. Future research should focus on improving mesh quality and ear detail through enhanced reconstruction and re-meshing, and on developing machine-learning-based refinement methods that use PR synthetic HRTFs as inputs and measured HRTFs as targets, with the aim of narrowing the remaining gap in spectral and perceptual fidelity.

\section{CherISH}

This research is carried out within the CherISH European Doctorate Network project, which aims to improve sound localisation skills for bilateral cochlear implants (CI) users through VR-based training. Spatial hearing rehabilitation using virtual reality headsets and headphones offers substantial practical advantages over traditional loudspeaker arrays, including accessibility, portability, and reduced infrastructure requirements \cite{parmar_virtual_2024}. However, the impact of individual spatial cues on training efficacy and rehabilitation outcomes for CI users remains unclear. The present work addresses this challenge by establishing an accessible framework for individual HRTF synthesis using minimal off-the-shelf hardware and available software tools. This pipeline provides a foundation for future investigations into spatial cue augmentation and personalised rehabilitation protocols for hearing-impaired populations.

\funding

Horizon-MSCA-2022-DN-01: CherISH is a European Doctorate Network project funded by the European Union's Horizon 2020 framework programme for research and innovation under the Marie Sklodowska-Curie Grant Agreement No: 101120054.

\conflict

The authors declare no conflicts of interest.

\ethics

This study was approved by the Research Governance and Integrity Team at Imperial College London. SETREC No. 7046527. Informed consent was obtained from all participants.

\dataavailability

The data and code used in this study will be made available upon reasonable request.

\end{document}